\documentclass[structabstract]{aa}  
\usepackage{graphicx}
\usepackage{txfonts}
\usepackage{subfigure}

\begin{document}
   \title{Comprehensive time series analysis of the transiting 
   extrasolar planet WASP-33b}

   \author{G.~Kov\'acs\inst{1,2}, T.~Kov\'acs\inst{1}, J.~D.~Hartman\inst{3},
           G.~\'A.~Bakos\inst{3,5}, 
           A.~Bieryla\inst{4}, D.~Latham\inst{4}, R.~W.~Noyes\inst{4}, Zs.~Reg\'aly\inst{1},
	       G.~A.~Esquerdo\inst{4}
          }

   \institute{Konkoly Observatory, Budapest, Hungary \\
              \email{kovacs@konkoly.hu}
              \and
	      Department of Physics and Astrophysics, University of North Dakota, Grand Forks, ND, USA 
              \and
              Department of Astrophysical Sciences, Princeton University, Princeton, NJ, USA
              \and
              Harvard-Smithsonian Center for Astrophysics, Cambridge, MA, USA
			\and
			Alfred P.~Sloan Research Fellow
             }

   \date{May 22, 2012; March 5, 2013}

 
  \abstract
   {HD 15082 (WASP-33) is the hottest and fastest rotating star known to
   harbor a transiting extrasolar planet  (WASP-33b).
   The 
   lack of high precision radial velocity (RV) data stresses the need 
   for precise light curve analysis and gathering further RV data.}
   {By using available photometric and RV data, we perform a blend 
   analysis, compute more accurate system parameters, confine the 
   planetary mass and attempt to cast light on the observed transit 
   anomalies.}
   {We combine the original HATNet observations and various followup data to jointly 
   analyze the signal content and extract the transit component and 
   use our RV data to aid the global parameter determination.}
   {The blend analysis of the combination of multicolor light curves 
    yields the first independent confirmation of the planetary 
    nature of WASP-33b. We clearly identify three frequency components 
    in the $15$--$21$~$d^{-1}$ regime with amplitudes $7$--$5$~mmag. 
    These frequencies correspond to the $\delta$~Scuti-type pulsation 
    of the host star. None of these pulsation frequencies or their 
    low-order linear combinations are in close resonance with the orbital 
    frequency. We show that these pulsation components explain some but 
    not all of the observed transit anomalies. The grand-averaged transit 
    light curve shows that there is a $\sim 1.5$~mmag brightening shortly 
    after the planet passes the mid-transit phase. Although the duration 
    and amplitude of this brightening varies, it is visible even through 
    the direct inspections of the individual transit events (some $40$--$60$\% 
    of the followup light curves show this phenomenon). We suggest that 
    the most likely explanation of this feature is the presence of a 
    well-populated spot belt which is highly inclined to the orbital 
    plane. This geometry is consistent with the inference from the 
    spectroscopic anomalies. Finally, we constrain the planetary mass 
    to $M_{\rm p}=3.27\pm0.73$~$M_{\rm J}$ by using our RV data 
    collected by the TRES spectrograph.} 
   {}


\titlerunning{Time series analysis of WASP-33}
\authorrunning{Kov\'acs et al.}
   \maketitle
%

%
%
\section{Introduction}

The short-period ($P=1.22$~d) transiting extrasolar planet 
WASP-33b was discovered by Christian et al.~(\cite{christian}) 
in the course of the transit survey conducted by the WASP 
project (Pollacco et al.~\cite{pollacco}). The host star, HD~15082,  
is a bright, $V=8.3$~mag A-type star with a high projected rotational 
velocity of $\sim 90$~km~s$^{-1}$. As a result, high precision radial 
velocity (RV) measurements usually demanded by the planet 
verification and planet mass estimation processes are very difficult 
to take. Followup spectroscopic observations by Collier Cameron 
et al.~(\cite{collier}, hereafter CC10) could only yield an upper limit of 
$4.1$~M$_{\rm J}$ on the planetary mass but no bisector analysis 
could be performed due to the several km~s$^{-1}$ scatter of 
the measured RV values. However, due to the large rotational 
velocity, the authors were able to utilize the surface velocity 
field mapping capability of the orbiting planet and showed very 
clearly that the orbital revolution was highly inclined 
(i.e., retrograde) in respect to the stellar rotational axis. 

The WASP-33 system stands out from the other transiting systems not 
only for its highest rotation rate and highest $T_{\rm eff}$ of 
$7400$~K (CC10) but also because of the planet's highest substellar 
temperature ($T_0=3800$~K, see Sect.~9). Due to the high incident 
flux and the early spectral type of the star we expect that a large 
amount of UV radiation is deposited in the planet's atmosphere. This, 
together with its high Roche-lobe filling factor (see Budaj \cite{budaj}) 
make this system perhaps the best candidate for studying effective 
planetary envelope evaporation and mass flow onto the host star. 

Due to its brightness and significant transit depth of $\sim 14$~mmag, 
the WASP-33 system is highly suitable for followup photometric 
observations. Some $80$\footnote{http://var2.astro.cz/ETD/} events 
have been observed over the years by professional and amateur 
astronomers. Most of these light curves (LCs) display various 
anomalies, such as transit depth variation, mid-transit humps, 
tilted full transit phase, asymmetric ingress/egress phases and 
small-amplitude oscillations. Although not all of these variations 
are necessarily real, it is clear that the system is peculiar and 
very much worth for further attention.  

WASP-33 was also observed by HATNet (Bakos et al., \cite{bakos02}, 
\cite{bakos04}) in the course of the search for transiting extrasolar 
planets (TEPs). After assigning the `candidate' status to the target in
January 2010, a followup reconnaissance spectroscopy conducted by the 
TRES spectrograph (F\H ur\'esz \cite{furesz}) has led to the conclusion 
that the rotational velocity was very high. Nevertheless, we continued 
the spectroscopic followup and found that the low velocity amplitude 
may suggest the presence of a planetary companion. After the SuperWASP 
announcement
we stopped pursuing this target but here we utilize both the RV and the 
photometric (HATNet and our early followup) data. 

In this paper we perform a comprehensive time series analysis by 
utilizing the HATNet data, the followup LCs deposited at the Exoplanet 
Transit Database (ETD, see Poddany, Brat, \& Pejcha (\cite{poddany})), 
other published photometric data and new LCs from the Fred Lawrence 
Whipple and Konkoly observatories.\footnote{Photometric data 
from HATNet, FLWO and Konkoly are available in electronic form at the 
CDS via anonymous ftp to cdsarc.u-strasbg.fr (130.79.128.5) or
via http://cdsweb.u-strasbg.fr/cgi-bin/qcat?J/A+A/volume/page. 
The linearly detrended light curves (as given in the Appendix) and 
the averaged light curve derived from them are also deposited 
at this site.}  
Our goal is to verify the planetary nature of WASP-33b purely from 
photometry (the `Kepler-way' of system validation -- see, e.g., 
Muirhead et al.~\cite{muirhead}), to derive more accurate system 
parameters and examine the possible causes of light curve anomalies. 
We also use our RV archive, based on the observations obtained by the 
TRES spectrograph, to improve the mass estimate of the planet.     

%
%
\begin{table*}[t!]
\caption{Photometric datasets on WASP-33}
\label{tab:datasets}
\footnotesize
\begin{center}
\begin{tabular}{lcrclclcrcl}
\noalign{\smallskip}
\hline\hline
\noalign{\smallskip}
LC\# & Date [UT]      &   N  & FIL & Source & &LC\# & Date [UT] & N & FIL & Source \\
\noalign{\smallskip}
\hline
\noalign{\smallskip}
01   & 2003\,--\,2007 & 2285 &  I  & HATNet      & &
23X  & 2011-10-13     &  505 &  J  & Kitt Peak      \\ 
02X  & 2010-03-03     &  275 &  z  & FLWO        & & 
24   & 2011-10-23     &   96 &  C  & LTremosa       \\ 
03   & 2010-08-23     &  245 &  R  & KHose       & &  
25   & 2011-11-12     &  291 &  I  & Konkoly        \\ 
04   & 2010-08-26     &  282 &  R  & FHormuth    & & 
26   & 2011-11-12     &   79 &  V  & GCorfini       \\ 
05   & 2010-08-26     &  163 &  R  & RNaves      & & 
27   & 2011-11-14     &  907 &  V  & SDvorak        \\ 
06   & 2010-09-17     &  220 &  R  & TScarmato   & & 
28X  & 2011-11-23     &  404 &  R  & JGaitan        \\
07   & 2010-09-28     &  277 &  R  & EHerrero    & & 
29X  & 2011-11-23     &  127 &  R  & RNaves         \\ 
08X  & 2010-10-11     &  598 &  B  & WDauberman  & & 
30   & 2011-11-24     &  446 &  I  & Konkoly        \\ 
09X  & 2010-10-15     &  191 &  R  & RNaves      & & 
31X  & 2011-11-24     &  177 &  V  & RNaves         \\
10   & 2010-10-20     &  434 &  R  & CLopresti   & & 
32   & 2011-12-01     &  888 &  I  & TDax           \\
11   & 2010-10-20     &  138 &  R  & RNaves      & & 
33   & 2011-12-10     &  146 &  C  & FGarcia        \\
12   & 2010-11-03     &  451 &  z' & Kitt Peak   & &
34   & 2011-12-15     &  109 &  I  & JGaitan        \\
13   & 2010-11-19     &  863 &  V  & SDvorak     & &
35X  & 2011-12-20     &  149 &  V  & DGorshanov     \\ 
14   & 2010-12-06     & 1254 &  C  & SShadick    & & 
36   & 2011-12-21     &  134 &  V  & RNaves         \\
15   & 2010-12-25     &  440 &  C  & AMaurice    & & 
37   & 2011-12-27     &  230 &  C  & JLopesino      \\ 
16   & 2011-01-05     &  437 &  C  & CWiedemair  & &  
38X  & 2012-01-01     &  113 &  V  & RNaves         \\
17   & 2011-01-13     &  778 &  V  & SDvorak     & & 
39X  & 2012-01-01     &  420 &  C  & JLopesino      \\
18   & 2011-01-16     &  238 &  C  & SGajdos     & &  
40   & 2012-01-09     &  316 &  I  & SShadick       \\ 
19   & 2011-09-21     &  311 &  I  & SShadick    & &  
41   & 2012-01-12     &  130 &  V  & RNaves         \\ 
20   & 2011-09-24     &  468 &  C  & LBrat       & &  
42   & 2012-01-20     &  534 &  I  & SShadick       \\ 
21X  & 2011-10-05     &  196 &  R  & RNaves      & & 
43   & 2012-02-11     &  365 &  I  & SShadick       \\ 
22   & 2011-10-13     &  310 &  \ H$_\alpha$ & Kitt Peak \\
\noalign{\smallskip}
\hline
\end{tabular}
\end{center}
\underline {Comments:} \\
{\em Header:} 
LC\#: light curve identification number used in this paper; 
Date: starting date of the observations;
N: number of data points (may differ from those given at the ETD site, e.g., because of multiple items); 
FIL: filter used in the observation (``C'' indicates that no filter was used);
Source: observer or observatory name. \\ 
{\em Notes:}
Plots of these light curves are presented in the main body of the paper and 
in the Appendix. 
Light curves labelled by ``X'' are not used in the analysis presented in this paper. 
They exhibit the following peculiarities. 
\#02: partial transit, difficulties in trend filtering; 
\#08: the transit is deep; 
\#09: deep and skewed transit;
\#21: deep and skewed transit;
\#23: near infrared ``J'' filter is used; 
\#28: large oscillations, earlier and narrower transit (the difference appears to be 28 min!);
\#29: large oscillations, same night observation as that of \#28 but no apparent severe shift in the transit center; 
\#31: deep and skewed transit, is in conflict with the regular behavior observed on the same night (see \#30); 
\#35: deeper by $>50$\% than the average transit; 
\#38: deep blip after the ingress;
\#39: deep blip after the ingress (the same night event as \#38).
\end{table*}
%


%
%
\section{HATNet detection}

WASP-33 is located in two HATNet fields, internally labeled as 
\#166 and 167. The observations were made through Bessel $I$ filter 
between November 06, 2003 and January 30, 2007 by the telescopes 
located at the Arizona (FLWO) and Hawaii (Mauna Kea) sites of HATNet. 
Altogether $12149$ data points have been gathered in three seasons, 
spanning $1181$~days. The data were collected with an integration 
time of $\sim 5$~min. We detected a robust transit signal by using 
the method of Box-fitting Least Squares (BLS, Kov\'acs, Zucker \& 
Mazeh \cite{kovacs02}) after applying the Trend Filtering Algorithm 
(TFA, Kov\'acs, Bakos \& Noyes \cite{kovacs05}) to remove systematics 
(note that the signal was detectable with high S/N without using TFA; 
however, filtering out systematics helped to increase S/N by a factor 
of two and substantially clean the folded light curve). 

\begin{figure}
   \centering
   \includegraphics[width=80mm]{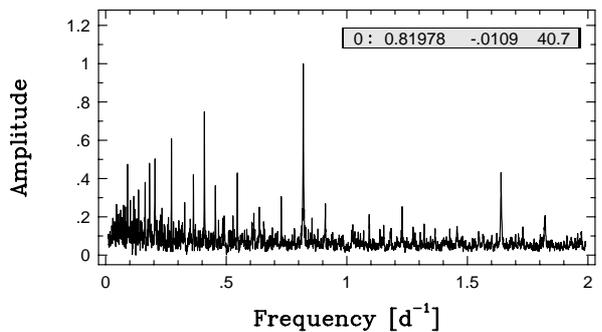}
      \caption{BLS spectrum of the TFA-filtered/reconstructed time series 
           based on the observations of WASP-33 by HATNet. The label in 
	   the upper right corner shows, from left to right, the 
	   pre-whitening order, peak frequency, transit depth 
	   (assuming trapezoidal transit shape) and the S/N of the 
	   peak frequency. The BLS amplitude is normalized to unity 
	   at the highest peak and refers to the signal without 
	   correction for blending by the nearby companion (see text).}
         \label{fig:hatnet-bls}
   \end{figure}

\begin{figure}
   \centering
   \includegraphics[width=80mm,angle=0]{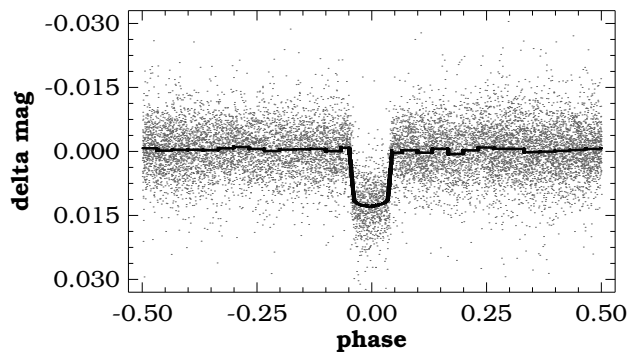}
      \caption{Folded light curve of the TFA-filtered/reconstructed time 
              series based on the observations of WASP-33 by HATNet 
	      (dots). The best-fitting simplified light curve model 
	      (with bin averaging in the out-of-transit part) 
	      is shown by the thick line.}
         \label{fig:hatnet-lc}
   \end{figure}

The BLS spectrum of the TFA-filtered data is shown in 
Fig.~\ref{fig:hatnet-bls}. The detection is very strong with the 
standard (sub)harmonics at integer frequency ratios. With the peak 
frequency of $0.819759$~$d^{-1}$ the signal was reconstructed 
(i.e., filtered out from systematics) by using a modified trapezoidal 
transit model with rounded bottom (to simulate limb darkening, 
with 15\% maximum deviation from the trapezoidal flat bottom) and 
an arbitrary out of transit variation (represented by $40$ bin 
averages of the light curve through the full orbital phase). The 
reconstructed light curve is plotted in Fig.~\ref{fig:hatnet-lc}. 
Except for the transit, we do not see any significant variation. We 
looked for an additional possible transit signal in the [0,2]~d$^{-1}$ 
frequency band in the pre-whitened signal by the above main transit. 
The result of this test was negative. A Fourier frequency analysis 
(or Discrete Fourier Transform -- DFT) in the same frequency band also 
concluded without finding a signal. The upper limit of any sinusoidal 
signal in the above frequency range is $0.5$~mmag. Both of these tests 
strongly suggest that we have a genuine transit signal at hand (see 
also the blend analysis in Sect.~4). 

The total transit duration (from the first to the last contact) 
is $2.7$~h, the ingress/egress (each) last for $14$~min. These values 
are in good agreement with the ones derived in other, accurate followup 
works (e.g., by CC10). However, the measured transit depth (computed 
with the above transit model) is $13$~mmag, which is in the lower tail 
of the distribution of the transit depths derived from the nine followup 
observations in the same color (see Table~\ref{tab:datasets} and Appendix). 
The average value of the depth from these data is $14$~mmag, with a 
standard deviation of $1.5$~mmag, much higher than the formal errors. If 
we consider only the most accurate data gathered by larger telescopes 
(i.e., Konkoly, CC10), we get a range of $13$-$17$~mmag, indicating that 
the transit depth might indeed vary. 

We note that the lower value of the transit depth derived from the HATNet 
data is not too likely to come from a third-light effect due to the 
close, $2.9$~mag fainter neighbor 2MASS~02265167+3732133. 
This object lies $1$~pix outside of the $2.4$~pix radius of the 
aperture used by the photometric code. With a PSF FWHM of $2.4$~pix 
some contamination is possible but this is most likely to be small.

%
%

%
%

%
%
\section{Followup datasets}

We use four photometric datasets in this paper: 
(i) the HATNet survey LC; (ii) the amateur/professional data deposited at 
the ETD site; (iii) the published followup LC of Herrero et al.~(\cite{herrero})  
and Sada et al.~(\cite{sada}); (iv) our own followup data obtained at the 
Whipple and Konkoly observatories. The list of all these datasets is 
given in Table~\ref{tab:datasets}. Several of them suffered from various 
trends due to instrumental or reduction effects. 
We attempted to correct these trends by fitting the following model 
light curve to the observations
%
%
\begin{equation}
LC_{\rm fit}(i) = c_1 + c_2TRAN(i) + c_3\Delta t(i) \hskip 1mm , 
\end{equation}
where $TRAN(i)$ is the simplified transit 
model\footnote{With the depth ratio of the rounded bottom kept 
constant for all colors.} described in Sect.~2, 
$\Delta t(i)=HJD(i)-HJD_{\rm mid}$, with $HJD_{\rm mid}$ denoting 
the middle of the observing time for the given followup observation, 
$c_1$, $c_2$ and $c_3$ are the regression coefficients to be determined 
by a standard least squares method, together with the transit parameters 
hidden in $TRAN(i)$. The light curves used in the analysis presented in 
this paper have been derived by subtracting the best fitting linear trend, 
as given by the coefficients $c_1$ and $c_3$. The detrended light curves, 
the transit parameters and the trend coefficient $c_3$ are presented in 
the Appendix (see Table~\ref{tab:transitpp}, Figures \ref{fig:etd1} and 
\ref{fig:etd2}). We also employed heliocentric corrections for most of 
the ETD LCs, as given at the ETD site. Since the HATNet data cover the 
full orbital phase, for the purpose of incorporating them in the analysis 
of the followup observations when constructing the average transit LC, 
we cut them in the phase interval $[-0.1,+0.1]$.  
 
The orbital period and center of transit times were updated through a BLS 
analysis of all available followup (non-flagged items in 
Table~\ref{tab:datasets}) and the HATNet data. The result is shown in 
Table~\ref{tab:periods}, where, for comparison, we also added the result 
derived from HATNet only. These values are in good agreement with those of 
CC10, albeit our period fits better the transit center of 
Herrero et al.~(\cite{herrero}). 
 
%
%
\begin{table}[t!]
\caption{Result of the BLS analysis}
\label{tab:periods}
\footnotesize
\begin{center}
\begin{tabular}{lrr}
\noalign{\smallskip}
\hline\hline
\noalign{\smallskip}
Data & Period [d]  &  $T_{\rm c}$ [HJD] \\
\noalign{\smallskip}
\hline
\noalign{\smallskip}
HATNet       &     $1.2198717$ &     $2452950.6718$ \\
             & $\pm 0.0000025$ &       $\pm 0.0019$ \\
HATNet$+$FUP &     $1.2198709$ &     $2452950.6724$ \\
             & $\pm 0.0000007$ &       $\pm 0.0015$ \\
\noalign{\smallskip}
\hline
\end{tabular}
\end{center}
\underline {Comments:} \\ 
Errors have been computed from Monte Carlo runs and correspond to $1\sigma$ limits. 
Number of data points, total time span and standard deviations of the simplified 
model fits [see text] are $12149$, $1181$, $6.1$~mmag and $24129$, $3019$ and $5.7$~mmag, 
for the HATNet and HATNet$+$FUP data, respectively.
\end{table}

\subsection{ETD data}

Due to its brightness, WASP-33 is high on the list of any followup 
programs, especially those conducted by amateur astronomers. As of 
June, 2012, there are $82$ entries listed at the ETD site on this 
object. Most of them have been collected by skilled amateur astronomers. 
In the analysis presented in this paper we use a large number of these 
followup data and omit only those that have especially high noise 
or very peculiar shape. In some cases these datasets are suspect to 
instrumental or environmental effects. Some of those that showed 
reasonably low noise level were flagged and left in 
Table~\ref{tab:datasets} for reference and further discussion in Sect.~6.  

\subsection{FLWO data}

As part of the HATNet follow-up program we obtained a light curve of
WASP-33 on the night of March 3/4, 2010 using the KeplerCam imager on
the FLWO~1.2\,m telescope and a Sloan~$z^{\prime}$ filter. These
observations began during ingress and ended before third contact. 
The images were calibrated and reduced to a light curve following 
the procedure discussed in Bakos et al.~(\cite{bakos10}).

\subsection{Konkoly data}

\begin{figure}
   \centering
   \includegraphics[width=80mm,angle=0]{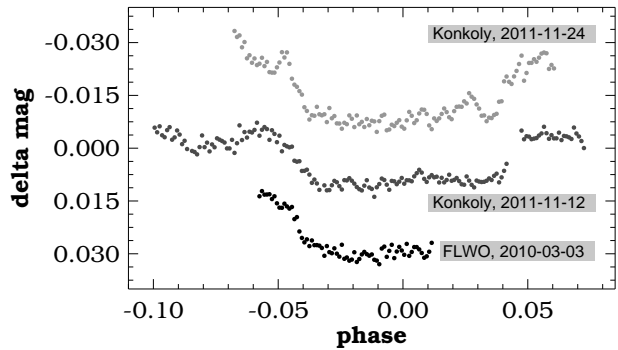}
      \caption{Photometric observations of WASP-33 at the Konkoly and Fred 
              Lawrence Whipple observatories. The Konkoly data have been 
	      taken through Cousins $I$, the FLWO data through Sloan $z$ 
	      filters. Data points are binned in $0.001$ phase (i.e., in 
	      105~s intervals).} 
         \label{fig:konkoly-flwo}
   \end{figure}

The transit was detected on November 12/13, 2011 by the $60/90$~cm 
Schmidt telescope located at Piszkestet\H o, Hungary. We gathered 
$2352$ exposures with $4$~sec integration time. The $4K\times4K$ 
images of the Apogee Alta CCD Camera were trimmed to $1K\times1K$ 
to speed up readout time. We used a filter in the Bessel $I$ band. 
Since the field is reasonably sparse around WASP33, we performed 
simple aperture photometry after the standard calibration procedure, 
including astrometry (P\'al \cite{pal2009}).  
Five comparison stars of similar brightness were used in a 
$\sim 0.3^{\circ}\times0.3^{\circ}\deg$ field of view centered on 
the target. After correcting for differential extinction (without the 
color term), we constructed the final light curve by averaging the data 
in $60$~s bins. As shown in Fig.~\ref{fig:konkoly-flwo}, the light curve 
exhibits both some oscillations (probably related to the $\delta$~Scuti-type 
pulsations of the host star) and a hump in the middle of the transit.

Yet another event was observed on the night of November 24/25, 2011 
with the Andor iXON 888 Electron Multiplying CCD (EMCCD) camera attached 
to the $50$~cm Cassegrain telescope at Piszkestet\H o. Altogether $12051$ 
exposures were taken with $\sim 1$~sec integration time, basically 
corresponding to the same sampling time due to the frame transfer mode. 
The field of view was $6'\times6'$ with a pixel resolution of $0.36$~arcseconds.  
As above, we used a Bessel filter in the $I$ band. 
After the standard calibration (dark removal and flat field correction) 
a special photometric data reduction was applied on the reduced images in 
the following steps (Regaly et al., in prep.): (i) $\sim30$ 
successive images were averaged after proper shifts due to atmospheric 
effects; (ii) the stellar fluxes on the co-averaged images were measured 
by an aperture photometric method in which the fuzzy apertures were 
determined by pixel-by-pixel S/N ratios; (iii) the individual stellar 
fluxes were corrected for airmass and by using only three reference 
stars -- the small field of view did not allow to use more -- we were 
able to reach an accuracy of $\sim 3\times10^{-3}$~mag on the above, 
$30$~sec-binned LC\@. The short-lasting glitches before the ingress and 
egress phases are attributed to amplified guiding errors due to the 
nearby faint companion.  

%
%

%
%
\section{Blend analysis}

CC10 previously validated the 
planetary nature of WASP-33b based on their Doppler tomography 
analysis of the system. They found that the Doppler tomography 
and transit light curve provide consistent values for the  
fractional area of WASP-33 that is eclipsed by WASP-33b, and 
concluded that there cannot be significant photometric dilution 
from a third body. To provide an independent validation of the 
planetary nature of WASP-33b, we conducted a blend analysis of 
the available light curves and absolute photometry following the 
procedure described in Hartman et al.~(\cite{hartman11}). (We recall 
that the traditional bisector span analysis does not work in the 
case of this star, due to its high rotational speed and the concomitant 
low accuracy of the radial velocity/bisector span values.)   

%
%
  \begin{figure}
   \centering
   \includegraphics[width=80mm,angle=0]{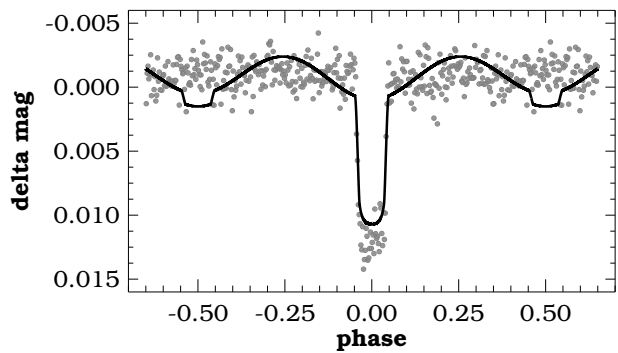}
      \caption{Folded/binned HATNet light curve of WASP-33 (dots) vs 
               the best-fitting blend model (continuous line). There are 
               400 bins throughout the full orbital period. The standard 
               deviations of the bin averages are very similar, they scatter 
               around $0.8$--$1.2$~mmag. The large out of transit variation 
	       of the presumed blended eclipsing binary excludes the possibility 
	       of a blend scenario by very high significance.}
         \label{fig:hat-blendanal}
   \end{figure}

We use the TFA-reconstructed HATNet LC and the summed-up followup 
LCs observed in $V$, $R$, $I$ and $z$ colors. (We did not use the 
non-specified filter-less ETD-posted measurements, their $B$ data and 
the $J$ and the $H_{\alpha}$ observations of Sada et al.~\cite{sada}).  
The HATNet data consists of $12149$ data points, whereas the followup 
observations in the four colors listed above, contribute by another $8361$ 
data points (with $2891$, $1759$, $3260$ and $451$ data points and $6$, 
$7$, $8$ and $1$ events in $V$, $R$, $I$ and $z$ colors, respectively). 
Concerning the accuracy of these averaged single color followup LCs in a 
$\sim 1$~min binning, we note that the bin averages have an overall 
standard deviation of $1$--$2$~mmag. (Slightly before/after the ingress/egress 
the scatter is higher, due to the sparser coverage of these parts by the 
followup observations.) Since the difference is small, for the blend 
analysis, we used all time series without filtering out by the $\delta$~Scuti 
components (see Sect.~5). In finding the best fitting blend model we used 
the observed LCs in all colors listed above. In this process the individual 
colors were treated separately and the best fit was defined as the best 
overall fit counting all colors. 

We find that a model consisting of a single star with a transiting planet 
is strongly preferred (with greater than $10\sigma$ confidence) over a 
blend between an eclipsing binary star system and another star 
(either foreground, or physically associated with the binary). The 
best-fit blend model shows both a significant out-of-transit variation 
($>3$\,mmag peak-to-peak) and a secondary eclipse ($\sim 1$\,mmag depth)
which are both ruled out by the HATNet light curve (see Sect.~2 and 
Fig.~\ref{fig:hat-blendanal}). 

This result enables us to derive more accurate system parameters by using 
the datasets employed in the blend analysis. We present the this global 
parameter determination in Sect.~8.   

%
%
\section{The $\delta$~Scuti pulsation components} 

It has been recognized quite early (see CC10)
that the host star of WASP-33b is also a variable 
(i.e., pulsating) star. The followup work by Herrero et al.~(\cite{herrero})  
clearly showed that WASP-33 was a $\delta$~Scuti-type star. This 
finding was later confirmed by Smith et al.~(\cite{smith}), 
Sada et al.~(\cite{sada}) and quite recently by Deming et al.~(\cite{deming}) 
in various infrared bands, involving both ground and space observations. 
The possibility of pulsations is not too surprising, since the physical 
parameters of the star fit well into the $\delta$~Scuti instability strip 
(although the relation between pulsation and physical parameters is not 
that simple in this part of the Hertzsprung-Russel diagram -- see 
Balona \& Dziembowski \cite{balona11b}, especially in the case of Am stars, 
such as the host of this system -- see CC10; Balona et al.~\cite{balona11a}). 

Herrero et al.~(\cite{herrero}) find that the frequency at 
$\sim 21.3$~d$^{-1}$ is in close (but high-order) resonance with 
the orbital frequency, which brings up the interesting possibility 
of tidally excited stellar oscillations by the planetary motion. 
In spite of this and other works mentioned above, so far, a clear 
detection and an accurate determination of the underlying frequency 
components is still missing. 

\begin{figure}
   \centering
   \includegraphics[width=80mm]{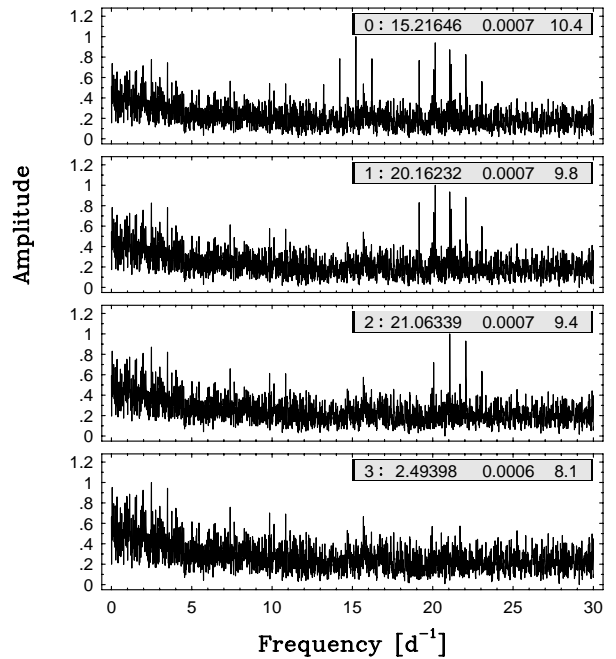}
      \caption{Successive pre-whitened DFT spectra of the HATNet data 
              after subtracting the transit component. The labels are 
	      defined in the same way as in Fig.~\ref{fig:hatnet-bls}}
         \label{fig:hatnet-dft}
   \end{figure}

We Fourier (DFT) analyzed the HATNet dataset after subtracting the 
transit signal. The analysis was performed in the $[0.0,30.0]$~d$^{-1}$ 
frequency band. The resulting spectra, shown in Fig.~\ref{fig:hatnet-dft}, 
clearly display the presence of three discrete sinusoidal components. 
The low (sub-mmag) amplitudes of all components and the relatively 
close frequency spacing of two of the components explain why the earlier 
attempts failed to identify all these components.\footnote{We checked 
also the available SuperWASP data (http://www.wasp.le.ac.uk/public/) but 
the noise level was too high to permit an in-depth analysis.} 

%
%

In employing these oscillations to clean up the followup light 
curves, we have to consider that the HATNet data were accumulated 
between 2003 and 2007 which implies that the frequency values derived 
solely from the HATNet data are not accurate enough to employ them 
for a useful prediction in the followup era (i.e., after 2010). 
Therefore, we filtered out the transit component from the followup 
data and analyzed the so-obtained residuals with the HATNet data 
(also without the transit signal).\footnote{In principle, due to the 
periodic light blocking effect of the transiting planet, we should deal 
with both variable stellar oscillations and transit shapes. However, due 
to the small size of the planet and the small amplitude of the pulsations, 
these are rather small effects, i.e., in the order of less than $2$\% on 
the pulsation amplitudes. We also disregard the possible color dependence 
of the pulsation amplitudes in the different filters (see, e.g., Liakos \& 
Niarchos~\cite{liakos}).} 
The Fourier decomposition resulted from this analysis is given in 
Table~\ref{tab:fourier-fit}.  

We may look at various linear combinations of these frequencies to see 
if Herrero's et al.~(\cite{herrero}) resonance hypothesis still holds. 
We find that with these well-resolved components there is no longer 
such a simple relation between the pulsation and orbital periods. The 
$\delta$~Scuti pulsation seems to be independent of the planetary 
companion. 

%
%
\begin{table}[t!]
\caption{Fourier decomposition of the $\delta$~Scuti pulsation of WASP-33}
\label{tab:fourier-fit}
\footnotesize
\begin{center}
\begin{tabular}{lcrrr}
\noalign{\smallskip}
\hline\hline
\noalign{\smallskip}
Data & i & $\nu_{\rm i}$ [d$^{-1}$] & $A_{\rm i}$ [mmag] & $\varphi_{\rm i}$ [rad]\\
\noalign{\smallskip}
\hline
\noalign{\smallskip}
HATNet        &1&    $15.2164306$ &     $0.758$ &     $3.982$ \\
              & & $\pm 0.0000409$ & $\pm 0.085$ & $\pm 0.218$ \\
              &2&    $20.1622967$ &     $0.733$ &     $5.087$ \\
              & & $\pm 0.0000497$ & $\pm 0.080$ & $\pm 0.244$ \\
              &3&    $21.0633928$ &     $0.719$ &     $4.442$ \\
              & & $\pm 0.0000485$ & $\pm 0.078$ & $\pm 0.225$ \\
	      
HATNet$+$FUP  &1&    $15.2151797$ &     $0.477$ &     $0.429$ \\
              & & $\pm 0.0000198$ & $\pm 0.054$ & $\pm 0.113$ \\
              &2&    $20.1623010$ &     $0.739$ &     $5.167$ \\
              & & $\pm 0.0000127$ & $\pm 0.053$ & $\pm 0.081$ \\
              &3&    $21.0634621$ &     $0.728$ &     $4.776$ \\
              & & $\pm 0.0000127$ & $\pm 0.049$ & $\pm 0.081$ \\
\noalign{\smallskip}
\hline
\end{tabular}
\end{center}
\underline {Comments:} \\ 
The following Fourier sum was fitted to the data (after pre-whitening 
by the transit signal): 
$A_0+\sum_{i=1}^3A_{\rm i}\sin(2\pi\nu_{\rm i}(t-t_0)+\varphi_{\rm i})$,  
where $t_0=2454500.0$. Errors have been computed from Monte Carlo runs 
and correspond to $1\sigma$ limits. 
\end{table}

\begin{figure}
   \centering
   \includegraphics[width=80mm,angle=0]{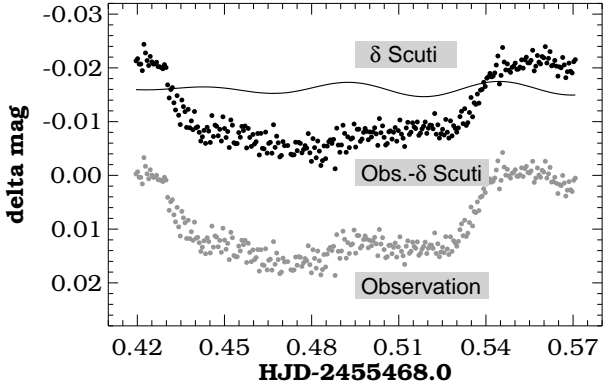}
      \caption{An example of the temporal coherence between the 
              observed oscillations and the ones predicted by 
	      the three high-frequency $\delta$~Scuti pulsation 
	      components (see Table~\ref{tab:fourier-fit}). The 
	      observations of Herrero et al.~(\cite{herrero}) 
	      have been plotted. Note that the amplitude 
	      of the pulsation is smaller than those of the observed 
              oscillations.}
         \label{fig:herrero-dsct}
   \end{figure}

We inspected the relation between the predicted $\delta$~Scuti 
pulsation and the oscillations observed/suspected in the individual 
followup LCs. The general conclusion is that we have a reasonable 
overall correlation between the two variations, although the 
observed amplitudes are larger and there are phase shifts 
occasionally. In Fig.~\ref{fig:herrero-dsct} we show an example 
on the stronger phase coherence and lack of complete ceasing of 
the transit anomaly after the subtraction of the $\delta$~Scuti 
components.       

%
%

%
%
\section{The mid-transit hump}

\begin{figure}
   \centering
   \includegraphics[width=80mm,angle=0]{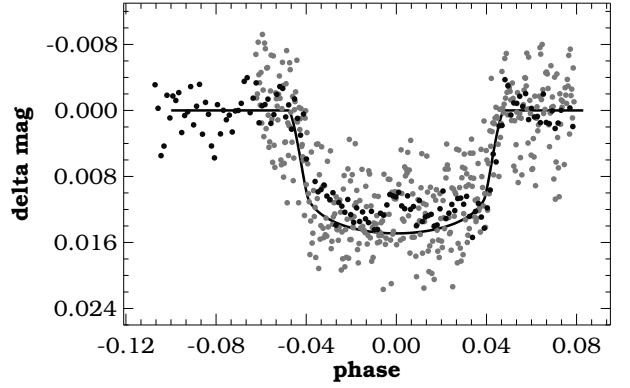}
      \caption{An example of the consistency of followup observations 
              made on the same date of 2010-10-20/21 by different 
	      observers. This is an example of a ``regular'' transit. 
	      Lighter dots are for the data of Claudio Lopresti (ETD), 
	      whereas the heavier ones for those of Ramon Naves (ETD). 
	      Continuous line shows the model fit to the grand-averaged  
	      transit light curve. The agreement between the two datasets 
	      is very good. Both indicate a small but significant 
	      brightening in the middle of the transit.}
         \label{fig:comp-08-35}
   \end{figure}

The $42$ followup light curves (in addition HATNet) listed in 
Table~\ref{tab:datasets} 
show a remarkable variety, with various oscillations and distortions, 
occasionally going to the extreme. Obviously, it is important to 
clarify if these changes are of astrophysical origin or just the result 
of unknown instrumental or environmental/weather-related effects. To get 
some information on the reality of certain specific features, we 
examine those LCs given in Table~\ref{tab:datasets} that were observed 
by different observers on the same nights. One such a night is 
2010-10-20/21. Fig.~\ref{fig:comp-08-35} shows that the two ETD 
observers Claudio Lopresti and Ramon Naves derived the same type 
of LCs, with a characteristic hump in the middle of the transit. 

%
%

\begin{figure}
   \centering
   \includegraphics[width=80mm,angle=0]{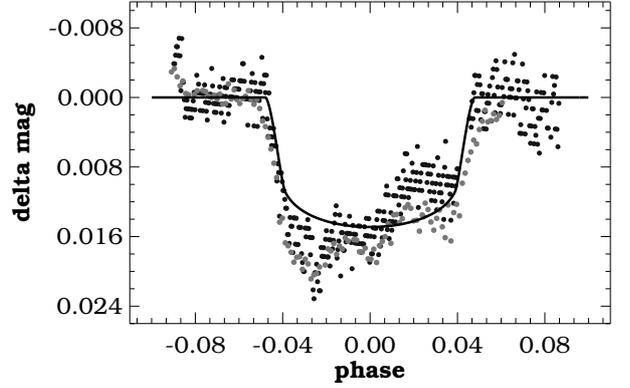}
      \caption{An example of the consistency of followup observations 
              made on the same date of 2012-01-01/02 by different 
	      observers. This is an example of a ``strongly discrepant'' 
	      transit. Lighter dots are for the data of Ramon Naves (ETD), 
	      whereas the heavier ones for those of Jordi Lopesino (ETD). 
	      Continuous line shows the model fit to the grand-averaged 
	      transit light curve. 
	      The agreement between the two datasets 
	      is very good which indicates that the large transit anomaly 
	      might be real.}
         \label{fig:comp-11-13}
   \end{figure}

In the above example the average shape of the LCs (except for the 
center hump) are in agreement with the one we expect from a standard 
transit. Vastly different (but consistent) LCs emerged from the 
observations made on 2012-01-01/02. The data come from two different 
sites and different telescopes from the ETD contributors Ramon Naves 
and Jordi Lopesino. The resulting LCs are plotted in 
Fig.~\ref{fig:comp-11-13}. The distortion is enormous but seems to be 
real from these independent datasets. 
(Nevertheless, as the referee noted, the relative proximity of 
the two telescopes used in these observations raises further suspicion 
on a possible weather-related issue of the phenomenon.)  

%
%

\begin{figure}
   \centering
   \includegraphics[width=80mm,angle=0]{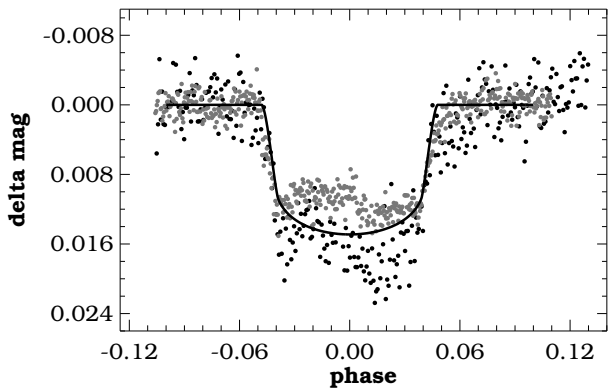}
      \caption{An example of the consistency of followup observations 
              made on the same date of 2011-10-13/14 through different 
	      filters (Sada et al.~\cite{sada}). This is an 
	      example of a ``mildly discrepant'' transit. Heavier  
	      dots are for the observations made through the 
	      H$_{\alpha}$, whereas the lighter ones for the $J$ 
	      filter. The larger scatter in the H$_{\alpha}$ band 
              is likely due to the smaller aperture of the telescope 
              used with this filter ($0.5$~m vs. $2.1$~m for the $J$ filter).  
	      Continuous line shows the model fit to the 
	      grand-averaged transit light curve. Both light curves 
	      clearly show the step-like variation of the flux with 
	      the break in the middle of the transit.}
         \label{fig:comp-39-40}
   \end{figure}

It is also worthwhile to compare observations made on the same 
night through different filters. Interestingly, the only available 
such datasets gathered by the Kitt Peak survey (Sada et al.~\cite{sada}) 
also show peculiarities which are consistent in the two filters 
(see Fig.~\ref{fig:comp-39-40}). We see in both colors a step-wise 
variation and a substantially shallower transit depth in color $J$. 
The difference is too large to be accounted for by a limb darkening 
effect.   

%
%

There are altogether seven parallel observations in the database 
shown in Table~\ref{tab:datasets}. They are all in very good agreement 
with each other (including ``amateur'' and ``professional'' data, i.e., 
the ones gathered on 2011-11-12/13), except for the event observed on 
2011-12-24/25 by Ramon Naves and Konkoly. While the Konkoly data (see 
Fig.~\ref{fig:konkoly-flwo}) display a regular transit shape, the 
parallel dataset shows a skewed transit (see Fig.~\ref{fig:etd2}). 
We think that this signals the need for substantial caution, especially 
in the case of excessive transit anomalies. 

There is one characteristic feature that appears in several LCs: 
we see a hump/brightening of a few mmag near the center of the transit. 
This anomaly has been noted also by Herrero et al.~(\cite{herrero}) 
and by Lubos Brat at the ETD site.\footnote{
The high-precision LC presented in a poster by Hardy \& Dhillon 
[in preparation, to be submitted to MNRAS; see 
www.ing.iac.es/astronomy/science/staff/posters/Liam\_Hardy\_poster.pdf] also 
shows a brightening near the transit center.} 
Since it seems to appear at the same orbital phase in each case, we decided 
to attempt a more definite detection of this feature by adding all 
non-discrepant LCs (i.e., the not labelled ones in Table~\ref{tab:datasets}) 
to increase the S/N. 

In summing up the LCs, one might be concerned about three effects 
influencing the final result: (i) the data were gathered in various 
filters that yield slightly different transit shapes due to the 
color dependence of the limb darkening; (ii) the LCs are of different 
quality; (iii) the phenomenon may not appear in all transits and then, 
averaging might decrease the S/N of the detection. 

If we had more accurate and less uniformly sampled data, concern 
(i) could be an important one. Due to limb darkening, the transit depth 
increases by $\sim 1.4$~mmag as we switch from filter $I$ to filter $B$. 
The difference shrinks to $\sim 1.0$~mmag if we compare filters $I$ and 
$V$. If we have roughly uniform coverage in the bins by all colors, then 
large part of the transit will be shifted in magnitude by roughly the 
same amount. This results in a conservation of the transit distortion, 
depending on the limb darkening very weakly. We did not see any improvement 
in the overall standard deviation of the bin averages when we scaled 
the transits by the expected depth dependence on color. Within this 
framework we also included the ``Clear'' ($C$, no-filter) data, although 
it is not clear without detailed consideration of the optical setting 
in each case what filter ``C'' really means. In selecting the reference 
model, we took the one corresponding to color ``B''. Various models 
corresponding to colors B--I gave increasing residual standard deviations 
toward redder colors (we fitted to the averaged LC by omitting the central 
hump). Redder colors tend to give more wavy residuals and worse fits  
to the ingress/egress parts but still a significant central hump 
(see Fig.~\ref{fig:ld-hump}).  

Concern (ii), the problem of different quality, was considered first 
within the standard framework of calculating the average of random variables 
of different variances. However, subsequent tests with various other ways of 
averaging have led to the conclusion of using the simplest way of averaging, 
with unit weights applied to all data points, no matter which LC are they 
attached to. We used the linear trend-filtered data (as derived in Sect.~3) 
in a binned simple summation of the non-flagged LCs in Table~\ref{tab:datasets}.  

Concern (iii) is very important, and, it seems to be justified on the basis 
of the current followup database (even if we consider that most of the 
followup data are too noisy to make statements on each individual events). 
For example, the Kitt Peak data (cf. Fig.~\ref{fig:comp-39-40}) clearly show 
that we have a stepwise anomaly rather than a slightly off-centered hump 
(at least on the night of the observations). The low-noise observations of 
Josep Gaitan (LC \#34 in Table~\ref{tab:datasets}) show apparently a flat 
bottom, whereas the LCs shown in Fig.~\ref{fig:comp-11-13} may support 
the possibility of large, asymmetric distortions (with the caveats mentioned 
above). Counterexamples 
on these varying anomalies are LCs \#11 and \#25 that show 
slight but steady humps that survive after pre-whitening by the $\delta$~Scuti 
components. We think that the noise level of the currently available 
data does not allow a clear selection of all LCs showing the specific 
distortion we are searching for. Therefore, the inclusion of all, 
non-peculiar LCs in the averaging process seems like a reasonable 
approach in the present situation.     

%
%
  \begin{figure}
   \centering
   \includegraphics[width=80mm,angle=0]{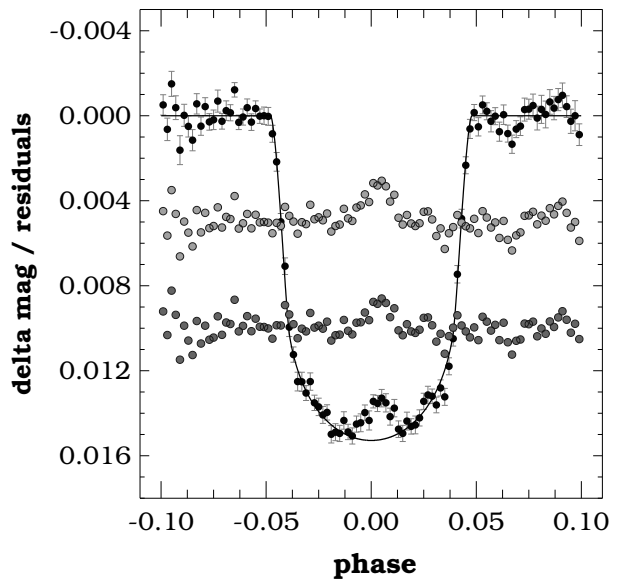}
      \caption{Transit light curve obtained by summing up 32 light 
              curves (see non-flagged items in Table~\ref{tab:datasets}). 
	      We combined the data in $0.002$ phase units. Error bars show 
	      the $\pm1\sigma$ ranges of the bin averages. Continuous line 
	      shows the reference model light curve (see text).  
	      The dots in the middle of the plot show the residuals 
	      (arbitrarily shifted) for the dataset without prewhitening 
	      by the $\delta$~Scuti components (light-shaded filled circles) 
	      and for the prewhitened one (dark-shaded filled circles).}
         \label{fig:grand_aver}
   \end{figure}

The result of the averaging process described above is displayed in 
Fig.~\ref{fig:grand_aver}. We see that the $\sim 1.5$~mmag hump is 
significant in this grand-averaged light curve. It is also visible that 
it is slightly off-centered. The hump is slightly less significant, 
if we subtract the $\delta$~Scuti component. Since the pulsation components 
are not in resonance with the orbital period, this effect of decreasing the 
hump amplitude is attributed to our inability to separate the two phenomena. 
This assumption is supported by the test we made with the predicted pulsation 
on the timebase of the followup data. Here we folded the predicted pulsational 
light curves by the orbital period and checked if the so-folded LC showed any 
structure, indicating a piling-up close to the transit center. We did not see 
such an effect, instead, we saw a roughly uniform distribution. On the other 
hand, there are cases when the observed wiggles do coincide with the 
$\delta$~Scuti component (see, e.g., Fig.~\ref{fig:herrero-dsct}). Therefore, 
a few such coincidences might already be sufficient to cause some decrease 
of the hump. 

Finally we note in passing that by adding additional followup data and/or 
using different mixtures of them -- but of course, by keeping the reliable 
ones -- do not result in a considerable change in the significance of the 
central hump. Following the recommendation of the referee, in the Appendix 
(see Fig.~\ref{fig:grand-aver3}) we show the grand-averaged LCs obtained 
from three different mixtures of followup data. Although with a somewhat 
lower significance, the anomaly is clearly exhibited in all average LCs, 
even though the new set of $13$ LCs are noisier by $\sim 20$\% on the average 
than the set presented in this paper.   

%
%
\section{Radial velocity data -- further constraining the planetary mass}

As a part of our regular routine on following up planetary candidates, 
spectral observations were taken with the Tillinghast Reflector Echelle 
Spectrograph (TRES, F\H ur\'esz \cite{furesz}) on the $1.5$~m Tillinghast 
Reflector at the Fred Lawrence Whipple Observatory in Arizona. Using 
the medium fiber, which has a resolving power of 
$\delta\lambda/\lambda\sim44000$ and a wavelength range of 
$3200$--$8900$~\r{A}, we obtained a total of $12$ observations taken 
between February 1, 2010 and October 25, 2010, to get complete phase 
coverage of our target. The spectra were extracted and analyzed using 
the procedures outlined in Buchhave et al.~(\cite{buchhave}). 
The radial velocities were measured by cross correlating a single spectral 
order (of the Mg~b triplet in $5150-5280$~\r{A}) against a synthetic 
template. Significant velocity variations were seen in the single order 
velocities but the scatter was fairly large, presumably due to the rapid 
rotation of this star. In an attempt to improve the accuracy, we calculated 
the multi-order velocities by cross correlating the spectra against each 
other order-by-order and summing the correlation functions. This 
increases the S/N by effectively reducing the measurement errors. We 
used the strongest observation as a template and covered the wavelength 
range of $\sim3980$--$5660$~\r{A}. The $12$ RV values together with 
their errors and spectral S/N values are listed in Table~\ref{tab:rv-tres}. 

%
%
\begin{table}[t]
\caption{Radial velocities measured by TRES on WASP-33}
\label{tab:rv-tres}
\footnotesize
\begin{center}
\begin{tabular}{crrr}
\noalign{\smallskip}
\hline\hline
\noalign{\smallskip}
BJD [UTC] & RV [$ms^{-1}$] & Error [$ms^{-1}$] & S/N \\
\noalign{\smallskip}
\hline
\noalign{\smallskip} 
2455228.602075  &     668.5  &     278.1  &     63.2 \\ 
2455237.594799  &     937.1  &     290.2  &     74.8 \\ 
2455239.621300  &     637.9  &     125.5  &    121.8 \\ 
2455240.577866  &       0.0  &     110.1  &    200.9 \\ 
2455241.574510  &     754.5  &     124.3  &    146.2 \\  
2455242.623591  &     880.4  &     137.8  &    151.5 \\ 
2455243.584301  &    1370.3  &     247.5  &    102.6 \\ 
2455243.610300  &    1383.3  &     110.1  &    167.2 \\  
2455251.664877  &    1044.5  &     509.6  &  (*)30.8 \\ 
2455252.642637  &     475.7  &     212.5  &     63.1 \\ 
2455461.002744  &     716.9  &     167.9  &     70.8 \\ 
2455494.798400  &     826.6  &     232.5  &    140.9 \\ 
\noalign{\smallskip}
\hline
\end{tabular}
\end{center}
\underline {Comments:} \\
The item denoted by (*) has been left out from the fit because of 
the outlier status, probably related to the low S/N. The velocity 
is on a {\em relative} scale. 
\end{table}

For consistency check with other spectroscopic/photometric results 
concerning the physical parameters of the star, we computed 
$T_{\rm eff}$, ${\rm v}_{\rm rot}\sin i$ 
and [M/H] by using the weighted 
average of the Mg~b triplet in three adjacent orders. Because the 
planetary nature of the system can be regarded as proven, we tamed 
the well-known ill-conditioning of the physical parameter determination 
by fixing the gravity $\log g$ as follows from the relation 
$g={4a/(\Delta t_{\rm tr})^2}$. (Here $a$ is the semi-major axis and 
$\Delta t_{\rm tr}$ is the transit duration and we assumed a circular 
orbit with small planet--star mass ratio -- see, e.g., Seager \& 
Mall\'en-Ornelas \cite{seager}; Sozzetti et al.~\cite{sozzetti2007}; 
and Beatty et al.~\cite{beatty}.) With $\log g=4.2$, for the strongest 
(highest S/N) spectra we got $T_{\rm eff}=7700\pm50$~K, 
${\rm v}_{\rm rot}\sin i=96\pm0.5$~kms$^{-1}$  
and [m/H]$=-0.10\pm0.08$. 
In a comparison with CC10, we see that our temperature and rotational 
velocity are higher by $270$~K and $10$~kms$^{-1}$, whereas the metallicity 
is lower by $0.2$~dex. Taking the formal errors given both by CC10 and us, 
these values differ by $2.4$, $19$ and $1$ sigma for $T_{\rm eff}$, 
${\rm v}_{\rm rot}\sin i$ and [M/H], respectively. Although the differences 
in $T_{\rm eff}$ and [M/H] are tolerable, we do not know what causes the 
formally highly significant discrepancy in ${\rm v}_{\rm rot}\sin i$. For 
this reason and for the overall better consistency of the temperature and 
metallicity values derived by CC10, we use their values in the determination 
of the system parameters in Sect.~8. 

\begin{figure}
   \centering
   \includegraphics[width=85mm]{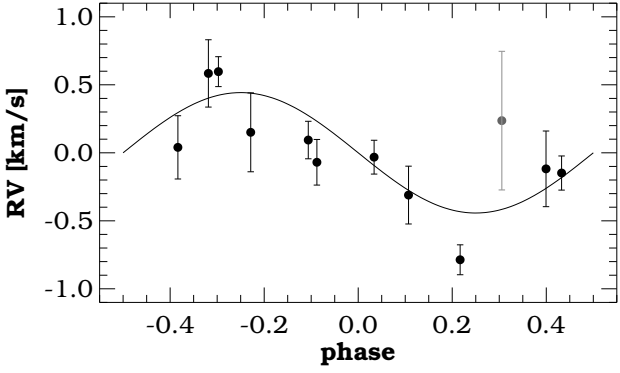}
      \caption{Phased radial velocity curve with the fitted sinusoidal 
               (assuming zero eccentricity). We fixed the period and 
	       the transit epoch as given by the joint analysis of the 
	       HATNet and followup data. Due to its large error, the 
	       gray-shaded data point at phase $\sim 0.3$ was left out 
	       of the fit.}
         \label{fig:tres-rv}
   \end{figure}

We fitted a sinusoidal signal (assuming zero eccentricity) to the 
$11$ RV values by omitting the point with the lowest S/N (which one 
is also an outlier in the fit). Since we fixed the period and the 
transit center according to the values obtained by the combined 
analysis of the HATNet and followup data, there remained two parameters 
to fit: (i) the RV semi-amplitude $K$ and (ii) the zero point shift. 
The resulting fit is shown in Fig.~\ref{fig:tres-rv}. Although the S/N 
of the fit is not particularly high, it is still possible to give a 
reasonable estimate on the RV amplitude and planetary mass. We get 
$K=0.443\pm0.095$~${\rm kms}^{-1}$ and, assuming a stellar mass and 
a semi-major axis of $1.495\pm0.030$~M$_{\rm sun}$ and 
$0.02556\pm0.0002$~AU respectively (see CC10), 
we get a planetary mass of 
$3.04\pm0.66$~M$_{\rm J}$.\footnote{Due to fixing the stellar and orbital 
parameters and their errors to those of CC10, 
the errors given here are smaller than the ones to be obtained in the global 
parameter fit in Sect.~8.} 
This is in agreement with the broad upper limit of $4.1$~M$_{\rm J}$ given 
by CC10 and also suggests that WASP-33b 
falls in the more massive class of planets. In Sect.~8 we get a more 
complete set of system parameters (including the planetary mass) by using a 
global fit to all observed quantities. 

%
%

%
%
\section{System parameters}

By using the RV data and the LCs employed in the blend analysis (in all three colors), 
we can derive more accurate system parameters. In this process we essentially followed  
the method described in our discovery papers (e.g., Bakos et al.~\cite{bakos11}). 
In short, we combine the parameter determination from the LC, RV and spectroscopic 
data through the intermediary of the stellar evolution isochrones (i.e., by using 
the Yale-Yonsei models of Yi et al.~\cite{yi2001}). The errors are estimated by a 
Markov Chain Monte Carlo method as an integral part of the parameter determination. 
Because we are using composite data, we did not perform trend filtering during the 
fit (we made a trend filtering as part of the data preparation -- see Sect.~3). 
One other difference from the Bakos et al.~(\cite{bakos11}) procedure is that we 
used a Mandel \& Agol (\cite{madel}) model for the HATNet data rather than a 
simplified flat-bottom transit model. To determine the physical parameters of the 
planet we used the atmospheric parameters ($T_{\rm eff}$, [Fe/H] and $\log g$) for 
WASP-33 taken from Collier Cameron et al.~(\cite{collier}) together with the 
constraint on the stellar density which comes from modeling the transit light curves. 
The orbital period and the moment of the transit center was fixed to the values we 
got by the BLS analysis of the full dataset (see Sect.~3). The eccentricity was 
also fixed to zero, as suggested by the infrared occultation data by 
Smith et al.~\cite{smith} and Deming et al.~\cite{deming}. Furthermore, because 
of the central hump, during the global analysis we omitted the data points in 
the phase interval $[-0.007,0.014]$. 

Table~\ref{tab:wasp33-params} lists the so-obtained parameters and their errors 
from the Monte Carlo method mentioned above. In comparing with the set derived by 
CC10, we see that the parameters are in good/fair agreement but our planet radius is 
$\sim 10$~\% larger. This difference comes mainly from the exclusion of the central 
hump region from our analysis (including the hump, the difference shrinks down to 
$\sim 4$~\%). Our errors, directly related to the light curve, are substantially lower.  
The absolute parameters (e.g., stellar mass) are of similar accuracy, or worse in 
our case. The reason for that is not entirely clear but might come from differences 
in the computational details and differences in the error limits. We agree also on 
the low age of the system, however our age estimate is limited to values larger 
than $0.1$~Gyr, to avoid confusion with pre-main sequence evolution tracks. However, 
it is worthwhile to note that systems with high temperature host stars are better 
candidates for accurate age estimates, due to the larger age spread in the same 
stellar density interval (see Fig.~5 of Collier Cameron et al.~\cite{collier}). 

%
%
\begin{table}[t]
\caption{System parameters of WASP-33}
\label{tab:wasp33-params}
\footnotesize
\begin{center}
\begin{tabular}{lll}
\noalign{\smallskip}
\hline\hline
\noalign{\smallskip}
Parameter & Value/Error \\
\noalign{\smallskip}
\hline
\noalign{\smallskip}
Transit LC parameters              &  \\
\hline
$R_{\rm p}/R_{\rm s}$              &  $0.1143\pm0.0002$ \\
$b$ (impact parameter)             &  $0.245\pm0.013$ \\
$T_{14}$ (total transit duration)  &  $0.1143\pm0.0002$~days \\
$T_{12}$ (ingress duration)        &  $0.0124\pm0.0002$~days \\  
RV curve parameters                &  \\
\hline
$K$ (RV semi-amplitude)            &  $463\pm102$~ms$^{-1}$ \\
RV fit RMS                         &  $251$~ms$^{-1}$ \\
Stellar parameters                 &  \\
\hline
$M_{\rm s}$                        &  $1.561_{-0.079}^{+0.045}$~$M_{\rm sun}$ \\
$R_{\rm s}$                        &  $1.509_{-0.027}^{+0.016}$~$R_{\rm sun}$ \\
$\log g_{\rm s}$                   &  $4.27\pm0.01$ [CGS] \\
$L_{\rm s}$                        &  $6.17\pm0.43$~$L_{\rm sun}$ \\  
$M_{\rm v}$                        &  $2.72\pm0.08$ [mag] \\
Age                                &  $0.4_{-0.2}^{+0.4}$~Gyr \\
Planet and orbital parameters      &  \\
\hline
$M_{\rm p}$                        &  $3.266\pm0.726$~$M_{\rm J}$ \\
$R_{\rm p}$                        &  $1.679_{-0.030}^{+0.019}$~$R_{\rm J}$ \\
$\log g_{\rm p}$                   &  $3.46_{-0.12}^{+0.08}$ [CGS] \\
$\rho_{\rm p}$                     &  $0.86\pm0.19$ [CGS] \\
$i$ (orbital inclination)          &  $86.2^{\circ}\pm0.2^{\circ}$ \\
$a$ (semi major axis)              &  $0.0259_{-0.0005}^{+0.0002}$~AU \\
$a/R_{\rm s}$                      &  $3.69\pm0.01$ \\
$d$ (distance)                     &  $117\pm2$~pc \\
\noalign{\smallskip}
\hline
\end{tabular}
\end{center}
\underline {Comments:}\\ 
Fixed parameters: $T_{\rm eff}=7430\pm100$, [M/H]$=0.1\pm0.2$, $P=1.2198709$~days, 
$T_{\rm c}=2452950.6724$ [HJD], $e=0$ (circular orbit is used, based on the occultation 
data of Smith et al.~\cite{smith} and Deming et al.~\cite{deming} -- see however 
de Mooij et al.~\cite{demooij}), $T_{12}=T_{34}$, i.e., ingress duration is 
assumed to be equal to the transit duration. The central hump phase of the 
light curve (see Fig.~\ref{fig:grand_aver}) was omitted in the global parameter 
determination. 
\\    
\end{table}

%
%
\section{Discussion}

To exhibit the special position of WASP-33b among the currently 
known TEPs, in Table~\ref{tab:roche-list} we summarize the 
relevant properties (from the point of view of this section) of 
the top 10 planets ordered by the value of the Roche-filling factor 
(Budaj \cite{budaj}). We see that WASP-33b is among the possible candidates 
for intensive star-planet interaction, including mass transfer. In the process 
of planet mass loss, the dynamical/gravitational distortion is coupled with 
the high temperature of the host star and the higher UV (continuum) flux 
impinging the planet surface facing the star. Considering that HD~209458 
(together with HD~189733b -- see Lecavelier des Etangs et al.~\cite{lecavelier}) 
has already been shown loosing mass (cf. Vidal-Madjar et al.~\cite{vidal}), 
and HD~209458 has a rather low Roche-filling factor, it is expected that 
those with much higher filling factors are very good candidates for 
observing mass-loosing planets. The high planet temperature, UV radiation 
and the bright apparent magnitude make WASP-33 especially a good target 
for observing mass loss.     

Although there are TEPs with strong stellar variability (e.g., 
CoRoT-Exo-4b, see Aigrain et al.~\cite{aigrain}), the transit 
anomalies detected in WASP-33 are unmatched. 
(Except for the extraordinary object KIC 12557548, supposedly a 
Mercury size ultra short period planet, showing dramatic changes in 
the transit depth and shape -- see Rappaport et al.~\cite{rappaport}.)  
The changes seen in the available followup data 
(even if we suspect that a considerable part of them may be of 
instrumental origin)  
are very large compared 
with the mild short-lasting humps attributed to stellar spot activity in 
some TEPs (e.g., CoRoT-2, HAT-P-11, WASP-4 and Kepler-17, see 
Nutzman et al.~\cite{nutzman}, Sanchis-Ojeda \& Winn~\cite{sanchis11a}, 
Sanchis-Ojeda et al.~\cite{sanchis11b}, and D\'esert et al.~\cite{desert}, 
respectively). Some of the anomalies observed are recurrent, 
appearing in similar phases in the transit (e.g., two spot belts 
in HAT-P-11, several of them in Kepler-17). This has led to the utilization 
of these quasi-stationary anomalies to estimate stellar obliquity, 
independently of the Rossiter-McLaughlin effect. This is very promising, 
although for a more reliable application of the method one needs 
well-covered, nearly continuous set of data, characteristic only of 
space observations. Even if this type of data are available, sufficiently 
long-lived, constant latitude spots are necessary to disentangle 
different geometric scenarii (e.g., HAT-P-11, see Sanchis-Ojeda \& Winn 
\cite{sanchis11a}).   

%
%
\begin{table}[t]
\caption{The first 10 TEPs ordered by the Roche-filling factor f(Roche)}
\label{tab:roche-list}
\footnotesize
\begin{center}
\begin{tabular}{lcccclr}
\noalign{\smallskip}
\hline\hline
\noalign{\smallskip}
Star      & f(Roche)  &  $T_0$ & P [d]  &   Age[Gyr] &  Sp  & V \\
\noalign{\smallskip}
\hline
\noalign{\smallskip}
WASP-12     &  0.613  & 3650 & 1.09  &  1.7 & G0  &   11.7\\
WASP-19     &  0.590  & 2900 & 0.79  &  0.6 & G8V &   12.3\\
WASP-4      &  0.425  & 2640 & 1.34  &  5.2 & G8  &   12.6\\
CoRoT-1     &  0.424  & 2680 & 1.51  &  7.2 & G0V &   13.6\\
OGLE-56     &  0.414  & 3180 & 1.21  &  2.0 & G   &   16.6\\
WASP-33     &  0.354  & 3800 & 1.22  &  0.1 & A5  &    8.3\\
WASP-17     &  0.333  & 2340 & 3.74  &  3.0 & F4  &   11.6\\
TrES-3      &  0.329  & 2340 & 1.31  &  0.3 & G   &   12.4\\
TrES-4      &  0.292  & 2500 & 3.55  &  2.9 & F   &   11.6\\
OGLE-132    &  0.288  & 2800 & 1.69  &  1.0 & F   &   16.8\\
HD 209458   &  0.229  & 2050 & 3.53  &  4.0 & G0V &    7.7\\
\noalign{\smallskip}
\hline
\end{tabular}
\end{center}
\underline {Comments:} \\
f(Roche) values are taken from Budaj (\cite{budaj}), ages and spectral types are 
from http://exoplanet.eu/, $T_0=T_{\rm eff}\sqrt{R_{\rm star}/a}$ is the 
equilibrium temperature at the substellar point (assuming circular orbit 
-- see Cowan \& Agol \cite{cowan}). The ages listed for WASP-4, CoRoT-1 and for 
TrES-3 are from Gillon et al.~\cite{gillon2009a}, \cite{gillon2009b} and from 
Sozzetti et al.~\cite{sozzetti2009}, respectively. For OGLE-132 we took an age 
1.0~Gyr as the average of the values given in Guillot et al.~(\cite{guillot}).  
\end{table}

Disregarding other light curve anomalies, since the hump near the center 
of the transit of WASP-33b seems to be recurrent, it is important to 
find an explanation on the physical origin of this feature. We 
investigated the following possibilities (in the order of likelihood). 

\begin{itemize}
\item[(a)]
Recurring spots (a belt of spots or a single spot that rotates 
resonantly with the orbital period)
\item[(b)]
Low spherical order distortions due to the $\delta$~Scuti pulsations
\item[(c)]
Variation of the projected planet size due to tidal distortion of the 
planet
\item[(d)]
Additional bodies (moons, planets)
\item[(e)]
Gravity darkening due to fast rotation
\end{itemize}

\paragraph{\bf Recurring spots}
Possibility (a) would have seemed to be a plausible explanation if we 
had forgotten about the spectral classification of the star. Metallic 
line A stars are supposed to be quiet objects with no or very weak 
magnetic fields, a prerequisite of spot activity. Furthermore, the 
element diffusion that causes the observability of the chemical 
peculiarities requires the absence of any mixing process, such as 
pulsation, if it excites random motion, i.e., turbulence. Nevertheless, 
as an important `by-product' of the SuperWASP survey, Smalley et al.~
(\cite{smalley}) have shown that nearly 13\% of the examined 1600 
stars classified as type Am are $\delta$~Scuti and $\gamma$~Dor stars 
pulsating with low amplitudes. Although it is not mentioned in their paper, 
some of the $\gamma$~Dor stars might be actually spotted stars with 
small spots. If this is true, then spots should not form a well-populated 
belt at a given latitude of activity, since in this case we could not 
observe a light variation. On the other hand, in the case of WASP-33 we 
{\em did not} detect a variation in the total flux above $0.5$~mmag 
in the $[0,2]$~d$^{-1}$ band, relevant for any rotational-related 
activities (see Sect.~2). Therefore, the spot belt assumption is more 
plausible. Then, this is consistent with the high tilt of the planetary 
orbit and the stellar rotational axis as suggested by the spectroscopic 
anomaly observed by Collier Cameron et al. (\cite{collier}). Interestingly, 
Kepler-17 also shows a strikingly similar central hump in the grand-averaged 
transit light curve (D\'esert et al.~\cite{desert}). While in the case 
of this system the hump can be explained by the presence of separate spots 
(since they cause observable effect in the average light variation), 
as mentioned above, no such a variation is seen in WASP-33. 

\paragraph{\bf Low spherical order distortions}
Explanation suggested in (b) sounds exciting, since it would explain 
in a natural way the steady-looking hump as a result of non-radial 
pulsation with $l=3$, $m=0$ spherical quantum numbers. The obvious 
condition for this model to work is the existence of a stroboscopic 
effect between the orbital and the pulsation frequencies. Since there 
is no such an effect (i.e., resonance), this model cannot produce 
a (quasi-)steady hump in the same orbital phase. 

\paragraph{\bf Variation of planet size}
Model (c) looks also a viable alternative for a moment but a glimpse 
on the possible distortion caused by the aspect dependence of a 
distorted planet in the short phase of transit (see Leconte, Lai 
\& Chabrier \cite{leconte}), clearly excludes this model from any 
further considerations as an explanation of the short-lasting central 
hump. 

\paragraph{\bf Additional bodies}
Possibility (d) is listed only for completeness, as an unlikely resort 
to a more exotic explanation of a steady transit anomaly. We would not 
only need an additional planet (or moon) on a resonant orbit but the 
stability condition would require this object to be of extreme low 
density. The observed hump of $\sim 1.5$~mmag implies an object radius 
of $\sim 0.6$~R$_{\rm J}$. With an ``ordinary'' gaseous planet density 
of $1$~gcm$^{-3}$ (i.e., the same as of WASP-33b), this results in a 
$\sim 0.2$~M$_{\rm J}$ object mass, 
which is certainly a reasonably high 
value to make the configuration unlikely to be stable. 

\paragraph{\bf Gravity darkening}
The observability of gravity darkening exerted by fast rotation (possibility 
(e)) has been brought up in the context of the high-precision photometric 
data attainable by the Kepler satellite (Barnes \cite{barnes}). It has 
apparently been observed in KOI-13, a short-period eclipsing binary with 
fast rotating components (v$_{\rm rot}\sim 70$~kms$^{-1}$) in the Kepler 
field (Szab\'o et al.~\cite{szabo}).  The amplitude of the observed anomaly 
is only $\sim 0.2$~mmag. Barnes' tests with a model system of an Altair 
($\alpha$~Aquilae)-type host star (with v$_{\rm rot}\sim 200$~kms$^{-1}$, 
e.g., van Belle et al.~\cite{van-belle}) indicate that the central hump 
might be as large as $1$~mmag in this case (see his Figs.~5 and 8). To 
extrapolate this result to the case of WASP-33, we can compare the ratios 
of the centrifugal and Newtonian components of the surface gravity 
${\rm v}_{\rm rot}^2R_{\rm star}/M_{\rm star}$. This yields an estimate of 
the fractional change of $T_{\rm eff}^4$ (in the classical von Zeipel 
approximation), the major contributor to the observed flux. We get that 
this ratio is about $6$-times larger for Barnes' model than for WASP-33. 
We conclude that model (e) is not preferred for WASP-33, since gravity 
darkening contributes probably at the level less than $\sim 0.2$~mmag to 
the distortion of the transit shape. 

As discussed in Sect.~6, there are anomalies in WASP-33 that are of much 
more substantial than the central hump. Even if several of these might come 
from observational errors, the explanation of the remaining ones (sloping  
full transit phase, different steepness of the ingress/egress phases, varying 
transit depth, asymmetric shifts in the ingress/egress phases) should imply 
major changes in the stellar flux. For example, the overall increase in the 
temporal transit depth 
(see notes added to Table~\ref{tab:datasets}) 
can only be explained if we assume a temporal but substantial dimming of the 
stellar light (or a sudden increase of the planet radius due to a violent 
Roche-lobe overflow). We do not have any idea on the explanation of the 
asymmetric distortions of the transit shape (but the case of KIC~12557548 is 
tempting -- see Rappaport et al.~\cite{rappaport}).
 
The issue of the variability of the nearby comparison stars as an explanation 
of the unusual transit anomalies has come up early in this work. 
The HATNet data allowed us to study the variability of the some $140$ 
objects in the trimmed field of the Konkoly Schmidt telescope (see Sect.~3). 
Since fainter stars contribute very little to the flux if only few 
comparison stars are used, we concentrated only on the 10 bright close 
neighbors employed by most of the followup works. Assuming that standard 
DFT and BLS analyses are the proper means to qualify their variability 
(i.e., no sudden -- non-periodic -- dimming or brightening occur in these 
stars), we found that none of them show the level of variability that 
could cause the observed strong anomalies. There are only two stars worth 
for further consideration, the rest do not show a significant variation 
above $0.5$--$1$~mmag. Star 2MASS~02263566+3735479 often appears as one 
of the comparison stars in the various followup works. This is a {\em variable}  
object with $f=3.12274$~d$^{-1}$ and a Fourier amplitude of $A=2.6$~mmag. 
The variability is slightly nonlinear (has a non-zero $2f$ component). The 
star is seldom used alone and even if it is done so, its variability introduces 
only some small, slightly nonlinear distortion (i.e., a trend) in the transit 
shape, that can be partially filtered out. Followup light curves using 
dominantly this comparison star do not show intriguing peculiarities. On the 
other hand, the very nearby bright (although still $2.9$~mag fainter) star 
2MASS~02265167+3732133 could be problematic if the aperture is not correctly 
set in the photometry. The HATNet data show that this 
object has some remaining systematics (i.e., $1$~d$^{-1}$ variation) at the 
$4$--$10$~mmag level. By judging from the accompanying maps at the ETD site, 
we think that most of the followup LCs were processed by small-enough apertures  
and this companion did not cause any important distortion in the 
target light curve. 
 
Finally, we draw attention to the extremely young age of the system (see Table 
\ref{tab:roche-list}). Although the early phase of planet formation, including 
the consumption of the gas component and the inward migration of the planet 
have probably ended, the debris disk might still be present 
(e.g., Krivov \cite{krivov}). This might still be observable in the far infrared. 
Furthermore, the retrograde orbit and the young age pose the intriguing question 
on the nature of the body that was able to perturb WASP-33b at such an early 
phase of evolution, right after the planet formation in a flat disk structure. 
The observations of Moya et al.~(\cite{moya}) by adaptive optics in the near 
infrared is a step in the direction of finding the perturber. If the low-luminosity 
object (a brown dwarf or a dwarf star) found by Moya et al.~(\cite{moya}) is/was 
gravitationally bounded to WASP-33, then perhaps a close violent encounter 
with WASP-33b at an early phase of planet formation in this stellar binary 
caused the near driving out of this planet by causing a perturbation that 
left it now on a retrograde orbit. Although planet-planet scattering due to 
dynamical instability in a plane-confined resonant multiplanetary system can 
also lead to highly inclined orbits (i.e., Libert \& Tsiganis \cite{libert}), 
this does not seem to be sufficient to excite retrograde orbits. 

%
%
\section{Conclusions}

WASP-33b is a peculiar planet both in term of its host star and in 
respect of the size of its tidal distortion and the radiation level 
received from its star. These properties, combined with its brightness, 
have made this target a very attractive one for various followup 
studies, in particular for photometric ones. 

Although our work primarily focused on the (time-domain) photometric 
properties of the system, by using our archival radial velocity (RV) 
data gathered by the TRES spectrograph we were able to give a better 
estimate the mass of WASP-33b. The derived mass of $3.27\pm0.73$~M$_{\rm J}$  
is in agreement with the earlier raw upper limit of $4.1$~M$_{\rm J}$ 
given by Collier Cameron et al.~(\cite{collier}). With the significant 
RV signal we can state that WASP-33b belongs to the more massive class 
of extrasolar planets.  

By using the HATNet database, we searched for the signature of 
short-period oscillations of the host star as reported earlier 
by some followup works (in particular, that of Herrero et al.~
\cite{herrero}). We confirmed the presence of $\delta$~Scuti-type 
pulsations, and, for the first time, we resolved the pulsation in 
three discrete components in the $15$--$21$~d$^{-1}$ range. All 
three peaks have sub-mmag amplitudes. Due to the well-known 
difficulties (mode identification, need for full evolutionary 
model structure, mode selection, etc. -- see, e.g., Casas et al.~
\cite{casas}) in using these components in a pulsation analysis 
to constrain the stellar parameters, we employed these oscillations 
only to test their effect on the finally derived grand-averaged transit 
light curve (which effect has proven to be negligible for the current data).  
Opposite to what was suggested 
by some earlier works (i.e., Herrero et al.~\cite{herrero}), nor 
the frequencies neither their linear combinations are in close 
resonance with the orbital period. 
 
Due to the lack of high precision RV data, we excluded possible blend 
scenarios primarily on the basis of the HATNet light curve. The 
absence of the predicted out of transit variation of the best-fitting 
blend model is a very strong support of the suspected genuine star--planet 
configuration based on the spectral tomography performed by Collier Cameron 
et al.~(\cite{collier}). As a ``by-product'' of the blend analysis we 
also derived more precise system parameters (see Table~\ref{tab:wasp33-params}).     
 
In addition to the high rotational speed and high temperature of 
the host star, the most remarkable peculiarity of this system is  
the numerous transit anomalies observed in the various followup 
light curves. In addition to the $\delta$~Scuti pulsations, these 
are extra features of the system. Interestingly, one type of 
anomalies, the {\em slightly off-centered $\sim 1$--$2$~mmag hump within 
the transit}, seems to persist in more than half of the followup data 
so that it becomes very nicely visible in the grand-averaged light 
curve, using $32$ followup light curves and the HATNet survey data. 
We attempted to give a phenomenological explanation of this 
feature by invoking, e.g., low-spherical order stellar pulsation, a 
belt-like starspot structure and variation in the projected cross 
section of a tidally distorted planet. We think that the star spot  
hypothesis could be a viable one but the existence of spots is 
questionable on the basis of the Am classification of the star. 

WASP-33 is a very interesting system for the substantial transit 
anomalies, for the retrograde orbit of the planet and for the young 
age of the parent star. The data used in the present work have been 
able to give a more accurate estimate of the basic system parameters 
and to show the quasi-permanent nature of the central small hump 
within the transit. However, because of the $\sim 0.1$\% effect, we were 
unable to show if this feature (perhaps with varying amplitudes) is 
present during all transit events. Also, because of the observational 
systematics and relatively high noise level, the reality of several 
(but not all) other anomalies can also be questioned. Therefore, 
gathering high-precision (sub-mmag/min) photometric transit data 
would be crucial in understanding this puzzling system.

\begin{acknowledgements}
The results presented in this paper have been greatly benefited 
from the archive of Exoplanet Transit Database (ETD) maintained 
by Stanislav Poddany, Lubos Brat and Ondrej Pejcha. We thank for 
their prompt and helpful feedback on our questions occurring during 
the analysis. We also thank to the numerous professional amateur 
observers who feed this site. Useful correspondence on the transit 
anomalies with ETD contributors Ramon Naves and Eugene Sokov are 
appreciated. Thanks are due to Pedro Valdes Sada, who sent us the 
data on WASP-33 gathered at Kitt Peak. We thank to Attila Szing for 
performing the observations on November 24/25, 2011 and to Andr\'as 
P\'al for making his astrometric routine available to us. Thanks are 
also due to Katalin Ol\'ah for discussions on stellar activity, to 
Jan Budaj for helping us to revisit the Roche-filling factor and to 
the referee whose thorough review helped us substantially in revising 
the paper. G.~K. and T.~K. acknowledge the support of the Hungarian 
Scientific Research Foundation (OTKA) through grant K-81373. G.~\'A.~B. 
and J.~D.~H. acknowledges partial support from NSF grant AST-1108686. 
Zs.~R. has been supported by the Hungarian OTKA Grants K-83790 and 
MB08C-81013, and the ``Lend\"ulet'' Program of the Hungarian Academy 
of Science. This research has made use of the SIMBAD database, 
operated at CDS, Strasbourg, France.  

\end{acknowledgements}

%
%
\newpage
\begin{appendix}
\section{Individual and various grand-averaged light curves, effect of limb 
darkening, transit and trend parameters}

%
%
  \begin{figure}
   \centering
   \includegraphics[width=63mm,angle=-90]{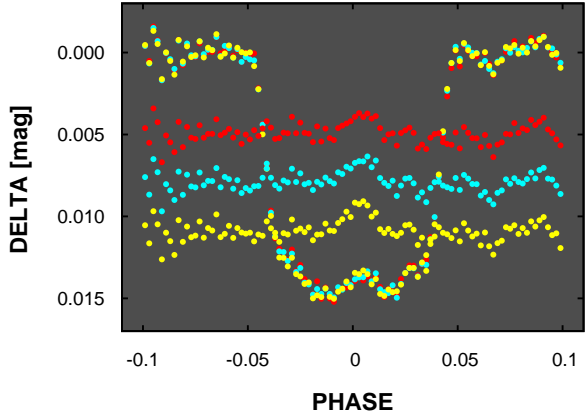}
      \caption{Grand-averaged transit light curves obtained from various 
               datasets. 
	       {\em Yellow:} our basic set of 32 LCs as presented in Sect.~6; 
	       {\em Cyan:} 32 LCs from the basic set and 13 LCs 
	       recommended by the referee from the ETD in her/his second 
	       report. [These LCs come from the following followups 
	       (name-date-filter): 
CLopresti-2010-09-11-R,  
SGajdos-2010-10-27-V, 
SKorotkiy-2010-10-25-R,  
JGarlitz-2010-11-19-R,    
LBrat-2010-11-27-C,        
Dhusar-2010-12-03-R,       
SShadick-2011-02-10-I,    
DSergison-2011-09-29-R,    
CGillier-2011-09-29-R,     
SShadick-2011-10-01-I,     
SShadick-2011-10-30-I,     
ABourdanov-2011-12-10-C,  
ERomas-2011-12-10-B.] 
              {\em Red:} as for the {\em cyan} set but omitting 11 stars 
	      recommended by the referee in her/his 3rd report. [These 
	      omitted LCs come from the following followups (name-date-filter):
SGajdos-2010-10-27-V,
SKorotkiy-2010-10-25-R,
LBrat-2010-11-27-C, and 
\#6, \#15, \#18, \#24, \#26, \#33, \#34, \#36 from our basic set as given 
              in Table~\ref{tab:datasets}.] 
              Notation and data sampling is the same as in Fig.~\ref{fig:grand_aver}. 
	      The residuals have been shifted vertically for better visibility.}
         \label{fig:grand-aver3}
   \end{figure}
%

%
%
  \begin{figure}
   \centering
   \includegraphics[width=63mm,angle=-90]{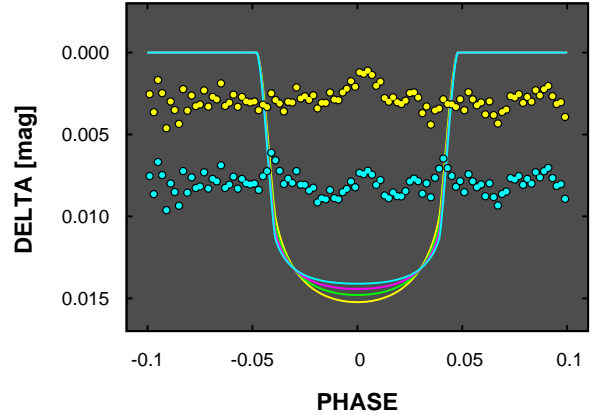}
      \caption{Effect of the assumed average color of the grand-averaged 
               light curve (Fig.~\ref{fig:grand_aver}) on the transit anomaly. 
	       The continuous curves show the best fitting models from $B$ 
	       (the deepest curve) through $V$, $R$ to $I$ (the shallowest 
	       curve). The dots represent the (shifted) residuals remaining 
	       after subtracting the model light curves corresponding to $B$ 
	       and $I$ (upper and lower sets of points, respectively). The 
	       ``kinks'' in the $I$ residuals in the ingress/egress parts 
	       suggest that shorter wavebands give a better representation 
	       of the mixture of colors the grand-averaged light curve is 
	       assembled from.}   
         \label{fig:ld-hump}
   \end{figure}
%

%
%
  \begin{figure*}
   \centering
   \includegraphics[width=160mm]{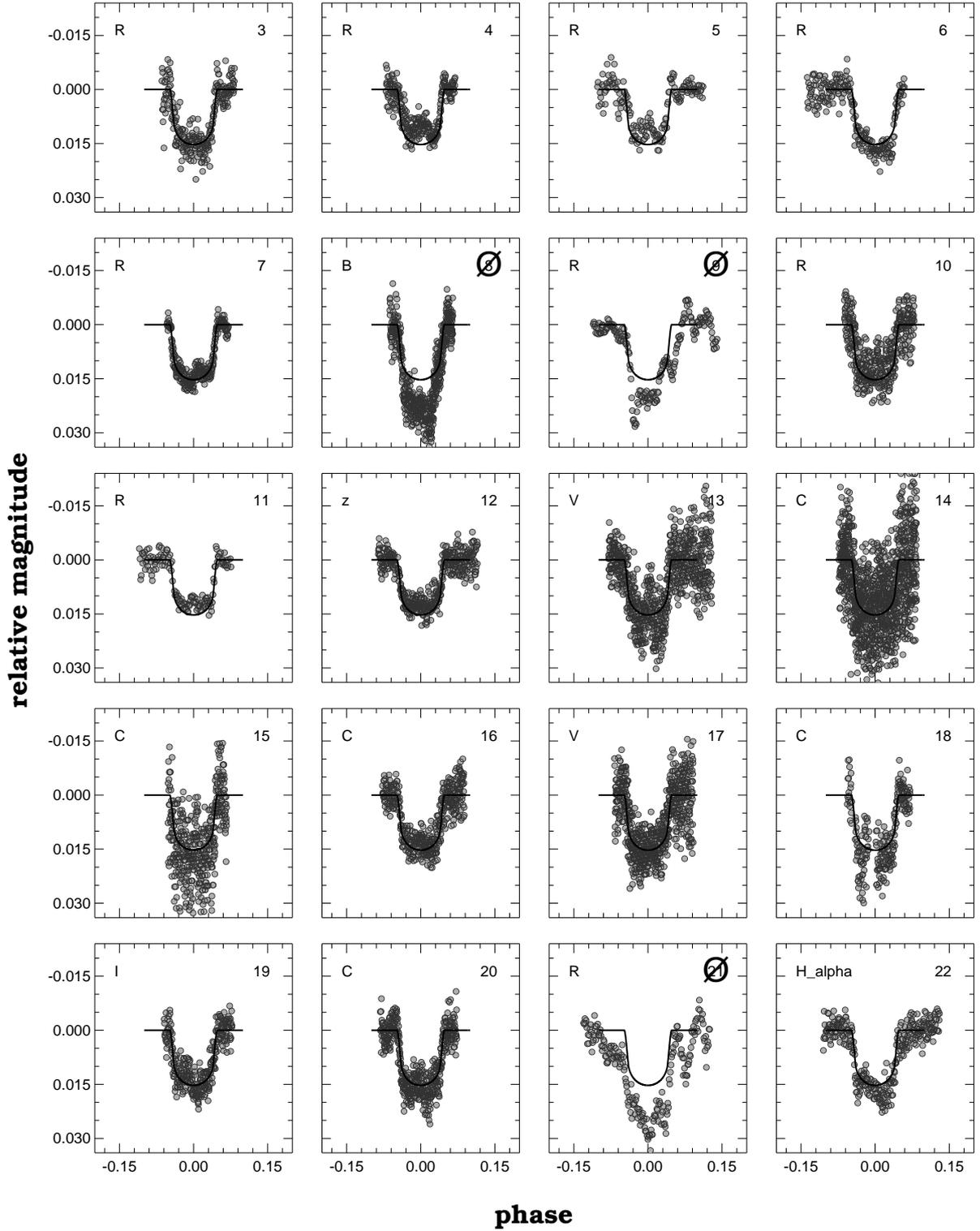}
      \caption{Light curves for the first half of the dataset as listed in 
               Table~\ref{tab:datasets}. Phase was computed from the ephemeris 
	       given for the full dataset in Table~\ref{tab:periods}. The 
	       light curves were linearly detrended as described in Sect.~3. 
	       The filters and the internal identification numbers (see 
	       Table~\ref{tab:datasets}) are given in each panel in the 
	       upper left and right corners, respectively (the sign 
	       $\varnothing$ is overplotted on the internal ID if the light 
	       curve is not included in the analysis presented in this paper). 
	       The reference model light curve is plotted by continuous line 
	       in each panel. Since the main purpose of the plot to display 
	       primarily the ETD light curves, for simplicity we left out 
	       \#01 (HATNet), \#02 (FLWO) and \#30 (Konkoly), shown in other 
	       figures in the paper.}
         \label{fig:etd1}
   \end{figure*}
%

%
%
  \begin{figure*}
   \centering
   \includegraphics[width=160mm]{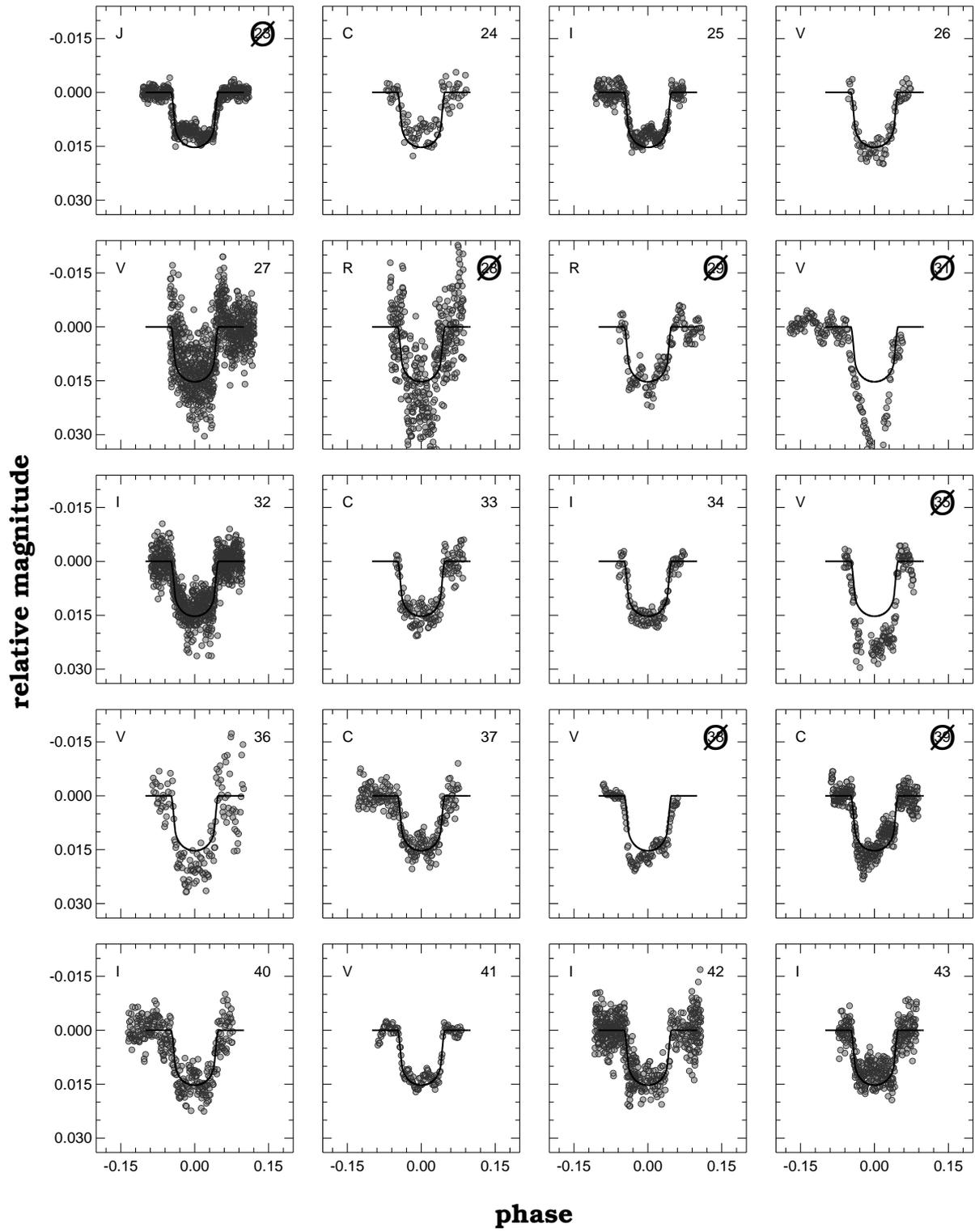}
      \caption{As in Fig.~\ref{fig:etd1} but for the second half of the 
               dataset.}
         \label{fig:etd2}
   \end{figure*}
%

%
%
\setcounter{table}{0}
\begin{table*}[t!]
\renewcommand\thetable{A-\arabic{table}}
\caption{Simplified transit model fits to the individual transit events 
of WASP-33}
\label{tab:transitpp}
\footnotesize
\begin{center}
\begin{tabular}{crrrrrrr}
\noalign{\smallskip}
\hline\hline
\noalign{\smallskip}
LC\# & Depth [mag] & $T12$ [d] & $T14$ [d] & $T_{\rm c}$ & $c_3$ & $\sigma_{\rm fit}$ &  N \\
\noalign{\smallskip}
\hline
\noalign{\smallskip}
01& $   -0.01292$ & $    0.00890$ & $    0.11098$ & $      52984.82950$ & $   -0.00022$ & $ 0.00458$ & $2196$\\
  & $\pm 0.00022$ & $\pm 0.00096$ & $\pm 0.00119$ & $\pm       0.00031$ & $\pm 0.00154$ & & \\
03& $   -0.01541$ & $    0.01497$ & $    0.11412$ & $      55431.89064$ & $   -0.01787$ & $ 0.00353$ & $ 245$\\
  & $\pm 0.00063$ & $\pm 0.00278$ & $\pm 0.00370$ & $\pm       0.00082$ & $\pm 0.00482$ & & \\
04& $   -0.01127$ & $    0.01059$ & $    0.10914$ & $      55435.54799$ & $   -0.02317$ & $ 0.00252$ & $ 282$\\
  & $\pm 0.00032$ & $\pm 0.00191$ & $\pm 0.00226$ & $\pm       0.00057$ & $\pm 0.00378$ & & \\
05& $   -0.01240$ & $    0.01356$ & $    0.10911$ & $      55435.55093$ & $    0.00652$ & $ 0.00268$ & $ 163$\\
  & $\pm 0.00057$ & $\pm 0.00418$ & $\pm 0.00505$ & $\pm       0.00095$ & $\pm 0.00298$ & & \\
06& $   -0.01632$ & $    0.01800$ & $    0.12099$ & $      55457.50954$ & $   -0.07299$ & $ 0.00262$ & $ 220$\\
  & $\pm 0.00062$ & $\pm 0.00282$ & $\pm 0.00340$ & $\pm       0.00072$ & $\pm 0.00402$ & & \\
07& $   -0.01437$ & $    0.01331$ & $    0.11643$ & $      55468.48488$ & $   -0.02304$ & $ 0.00174$ & $ 277$\\
  & $\pm 0.00035$ & $\pm 0.00111$ & $\pm 0.00138$ & $\pm       0.00035$ & $\pm 0.00339$ & & \\
08& $   -0.02460$ & $    0.03090$ & $    0.12431$ & $      55480.68451$ & $    0.08107$ & $ 0.00382$ & $ 598$\\
  & $\pm 0.00040$ & $\pm 0.00103$ & $\pm 0.00080$ & $\pm       0.00051$ & $\pm 0.00488$ & & \\
09& $   -0.02051$ & $    0.05230$ & $    0.16881$ & $      55485.54744$ & $    0.04160$ & $ 0.00358$ & $ 191$\\
  & $\pm 0.00098$ & $\pm 0.00549$ & $\pm 0.00626$ & $\pm       0.00147$ & $\pm 0.00416$ & & \\
10& $   -0.01284$ & $    0.00878$ & $    0.10884$ & $     55490.44146$ & $     0.03935$ & $ 0.00428$ & $  434$\\
  & $\pm 0.00053$ & $\pm 0.00254$ & $\pm 0.00319$ & $\pm      0.00064$ & $\pm  0.00499$ & & \\
11& $   -0.01235$ & $    0.00973$ & $    0.11151$ & $      55490.44463$ & $   -0.02676$ & $ 0.00185$ & $ 138$\\
  & $\pm 0.00040$ & $\pm 0.00278$ & $\pm 0.00306$ & $\pm       0.00073$ & $\pm 0.00243$ & & \\
12& $   -0.01265$ & $    0.01494$ & $    0.11766$ & $     55503.86355$ & $    -0.01721$ & $ 0.00240$ & $  451$\\
  & $\pm 0.00029$ & $\pm 0.00191$ & $\pm 0.00210$ & $\pm      0.00045$ & $\pm  0.00199$ & & \\
13& $   -0.01575$ & $    0.01336$ & $    0.11250$ & $     55519.71991$ & $     0.02985$ & $ 0.00568$ & $  863$\\
  & $\pm 0.00049$ & $\pm 0.00278$ & $\pm 0.00315$ & $\pm      0.00059$ & $\pm  0.00290$ & & \\
14& $   -0.01425$ & $    0.01566$ & $    0.12709$ & $     55536.79915$ & $     0.09255$ & $ 0.00909$ & $ 1254$\\
  & $\pm 0.00061$ & $\pm 0.00452$ & $\pm 0.00512$ & $\pm      0.00104$ & $\pm  0.00558$ & & \\
15& $   -0.01695$ & $    0.01077$ & $    0.11811$ & $     55556.31457$ & $     0.01598$ & $ 0.00745$ & $  440$\\
  & $\pm 0.00115$ & $\pm 0.00370$ & $\pm 0.00393$ & $\pm      0.00094$ & $\pm  0.01329$ & & \\
16& $   -0.01425$ & $    0.01666$ & $    0.11604$ & $      55567.29575$ & $   -0.01508$ & $ 0.00325$ & $ 437$\\
  & $\pm 0.00035$ & $\pm 0.00201$ & $\pm 0.00264$ & $\pm       0.00070$ & $\pm 0.00376$ & & \\
17& $   -0.01524$ & $    0.01272$ & $    0.10953$ & $      55574.61261$ & $    0.02431$ & $ 0.00518$ & $ 778$\\
  & $\pm 0.00046$ & $\pm 0.00222$ & $\pm 0.00247$ & $\pm       0.00059$ & $\pm 0.00365$ & & \\
18& $   -0.01816$ & $    0.01187$ & $    0.11168$ & $     55578.27423$ & $     0.06357$ & $ 0.00455$ & $  238$\\
  & $\pm 0.00088$ & $\pm 0.00374$ & $\pm 0.00449$ & $\pm      0.00088$ & $\pm  0.00864$ & & \\
19& $   -0.01453$ & $    0.01675$ & $    0.11957$ & $      55825.90836$ & $   -0.03323$ & $ 0.00315$ & $ 311$\\
  & $\pm 0.00057$ & $\pm 0.00247$ & $\pm 0.00285$ & $\pm       0.00073$ & $\pm 0.00599$ & & \\
20& $   -0.01604$ & $    0.01095$ & $    0.11408$ & $      55829.56788$ & $    0.00009$ & $ 0.00326$ & $ 468$\\
  & $\pm 0.00034$ & $\pm 0.00175$ & $\pm 0.00183$ & $\pm       0.00038$ & $\pm 0.00298$ & & \\
21& $   -0.02118$ & $    0.01743$ & $    0.12884$ & $      55840.54767$ & $    0.06557$ & $ 0.00407$ & $ 196$\\
  & $\pm 0.00073$ & $\pm 0.00383$ & $\pm 0.00480$ & $\pm       0.00102$ & $\pm 0.00430$ & & \\
22& $   -0.01657$ & $    0.01132$ & $    0.11590$ & $     55847.86651$ & $     0.04421$ & $ 0.00287$ & $  310$\\
  & $\pm 0.00047$ & $\pm 0.00173$ & $\pm 0.00205$ & $\pm      0.00051$ & $\pm  0.00241$ & & \\
23& $   -0.01159$ & $    0.01211$ & $    0.12316$ & $     55847.86687$ & $     0.01058$ & $ 0.00136$ & $  505$\\
  & $\pm 0.00015$ & $\pm 0.00093$ & $\pm 0.00098$ & $\pm      0.00027$ & $\pm  0.00073$ & & \\
24& $   -0.01131$ & $    0.00366$ & $    0.10183$ & $     55857.62251$ & $    -0.03481$ & $ 0.00258$ & $   96$\\
  & $\pm 0.00059$ & $\pm 0.00303$ & $\pm 0.00370$ & $\pm      0.00077$ & $\pm  0.00548$ & & \\
25& $   -0.01273$ & $    0.01525$ & $    0.11739$ & $      55878.36071$ & $   -0.01035$ & $ 0.00174$ & $ 291$\\
  & $\pm 0.00025$ & $\pm 0.00173$ & $\pm 0.00224$ & $\pm       0.00046$ & $\pm 0.00217$ & & \\
26& $   -0.01605$ & $    0.01424$ & $    0.11138$ & $      55878.36171$ & $    0.03985$ & $ 0.00223$ & $  79$\\
  & $\pm 0.00080$ & $\pm 0.00301$ & $\pm 0.00373$ & $\pm       0.00106$ & $\pm 0.00915$ & & \\
27& $   -0.01390$ & $    0.01724$ & $    0.10546$ & $      55879.58875$ & $   -0.04512$ & $ 0.00620$ & $ 907$\\
  & $\pm 0.00071$ & $\pm 0.00278$ & $\pm 0.00383$ & $\pm       0.00092$ & $\pm 0.00504$ & & \\
28& $   -0.02336$ & $    0.01086$ & $    0.08862$ & $      55889.32128$ & $     0.05757$ & $ 0.00921$ & $  404$\\
  & $\pm 0.00122$ & $\pm 0.00411$ & $\pm 0.00531$ & $\pm       0.00070$ & $\pm  0.01072$ & & \\
29& $   -0.01469$ & $    0.00985$ & $    0.11435$ & $      55889.33912$ & $    0.03363$ & $ 0.00305$ & $ 127$\\
  & $\pm 0.00076$ & $\pm 0.00320$ & $\pm 0.00386$ & $\pm       0.00090$ & $\pm 0.00607$ & & \\
\noalign{\smallskip}
\hline
\end{tabular}
\end{center}
\underline {Comments:} \\ 
See Sect.~2 for the description of the simplified transit model. 
{\em Header:} 
LC\#: light curve identifying number (see Table~1); 
Depth: transit depth;
$T12$: ingress(=egress) duration; 
$T14$: total transit duration; 
$T_{\rm c}$: time of the center of transit; 
$c_3$: linear trend coefficient (see Eq. (1); negative values indicate 
overall brightening);  
$\sigma_{\rm fit}$: standard deviation of the residuals throughout the 
total time span of the followup data;
N: number of data points. 
{\em Note:} The partial transit observation from FLWO (\#02) has been left out, due to 
numerical difficulties in fitting the trend and the transit models simultaneously. The 
center of transit time of \#28 seems to be strongly influenced by the large oscillations, 
leading to an estimate of $28$~ min earlier than predicted. 
\end{table*}
%
%
\setcounter{table}{0}
\begin{table*}[t!]
\renewcommand\thetable{A-\arabic{table}}
\caption{Continued}
\label{tab:transitpp}
\footnotesize
\begin{center}
\begin{tabular}{crrrrrrr}
\noalign{\smallskip}
\hline\hline
\noalign{\smallskip}
LC\# & Depth [mag] & $T12$ [d] & $T14$ [d] & $T_{\rm c}$ & $c_3$ & $\sigma_{\rm fit}$ &  N \\
\noalign{\smallskip}
\hline
\noalign{\smallskip}
30& $   -0.01670$ & $    0.01081$ & $    0.11132$ & $      55890.56016$ & $   -0.01488$ & $ 0.00302$ & $ 446$\\
  & $\pm 0.00034$ & $\pm 0.00111$ & $\pm 0.00138$ & $\pm       0.00034$ & $\pm 0.00391$ & & \\
31& $   -0.03987$ & $    0.05448$ & $    0.13802$ & $      55890.57094$ & $    0.05401$ & $ 0.00331$ & $ 177$\\
  & $\pm 0.00258$ & $\pm 0.00399$ & $\pm 0.00223$ & $\pm       0.00069$ & $\pm 0.00492$ & & \\
32& $   -0.01367$ & $    0.01229$ & $    0.11489$ & $     55896.66071$ & $    -0.04361$ & $ 0.00355$ & $  888$\\
  & $\pm 0.00031$ & $\pm 0.00163$ & $\pm 0.00176$ & $\pm      0.00041$ & $\pm  0.00187$ & & \\
33& $   -0.01518$ & $    0.00906$ & $    0.11289$ & $      55906.41511$ & $    0.08610$ & $ 0.00279$ & $ 146$\\
  & $\pm 0.00071$ & $\pm 0.00204$ & $\pm 0.00238$ & $\pm       0.00061$ & $\pm 0.00789$ & & \\
34 & $   -0.01587$ & $    0.01721$ & $    0.11740$ & $      55911.29675$ & $    0.00249$ & $ 0.00172$ & $ 109$\\
  & $\pm 0.00044$ & $\pm 0.00194$ & $\pm 0.00234$ & $\pm       0.00059$ & $\pm 0.00394$ & & \\
35& $   -0.02361$ & $    0.01385$ & $    0.12109$ & $      55916.17801$ & $    0.00774$ & $ 0.00264$ & $ 149$\\
  & $\pm 0.00062$ & $\pm 0.00132$ & $\pm 0.00148$ & $\pm       0.00044$ & $\pm 0.00580$ & & \\
36& $   -0.01990$ & $    0.00574$ & $    0.10720$ & $      55917.39612$ & $    0.13020$ & $ 0.00566$ & $ 134$\\
  & $\pm 0.00105$ & $\pm 0.00335$ & $\pm 0.00410$ & $\pm       0.00072$ & $\pm 0.00722$ & & \\
37& $   -0.01489$ & $    0.03037$ & $    0.13544$ & $      55923.49690$ & $    0.05587$ & $ 0.00283$ & $ 230$\\
  & $\pm 0.00096$ & $\pm 0.01478$ & $\pm 0.02050$ & $\pm       0.00148$ & $\pm 0.00472$ & & \\
38& $   -0.01680$ & $    0.01616$ & $    0.12705$ & $      55928.37473$ & $   -0.02463$ & $ 0.00200$ & $ 113$\\
  & $\pm 0.00055$ & $\pm 0.00220$ & $\pm 0.00289$ & $\pm       0.00060$ & $\pm 0.00517$ & & \\ 
39& $   -0.01468$ & $    0.01312$ & $    0.11406$ & $      55928.37224$ & $   -0.02589$ & $ 0.00299$ & $ 420$\\
  & $\pm 0.00033$ & $\pm 0.00233$ & $\pm 0.00300$ & $\pm       0.00057$ & $\pm 0.00258$ & & \\
40& $   -0.01541$ & $    0.01615$ & $    0.12301$ & $      55935.69728$ & $   -0.00475$ & $ 0.00356$ & $ 316$\\
  & $\pm 0.00065$ & $\pm 0.00287$ & $\pm 0.00316$ & $\pm       0.00071$ & $\pm 0.00398$ & & \\
41& $   -0.01402$ & $    0.01076$ & $    0.11294$ & $      55939.35322$ & $    0.02748$ & $ 0.00150$ & $ 130$\\
  & $\pm 0.00032$ & $\pm 0.00102$ & $\pm 0.00137$ & $\pm       0.00038$ & $\pm 0.00219$ & & \\
42& $   -0.01384$ & $    0.00736$ & $    0.11353$ & $     55946.67657$ & $     0.04735$ & $ 0.00414$ & $  534$\\
  & $\pm 0.00043$ & $\pm 0.00245$ & $\pm 0.00246$ & $\pm      0.00068$ & $\pm  0.00233$ & & \\
43& $   -0.01245$ & $    0.01017$ & $    0.11581$ & $     55968.63311$ & $     0.05396$ & $ 0.00300$ & $  365$\\
  & $\pm 0.00040$ & $\pm 0.00169$ & $\pm 0.00193$ & $\pm      0.00054$ & $\pm  0.00274$ & & \\
\noalign{\smallskip}
\hline
\end{tabular}
\end{center}
\end{table*}

\end{appendix}

\end{document}